\documentclass[aps,twocolumn,prd,superscriptaddress,preprintnumbers,showpacs]{revtex4}
\usepackage{graphicx}
\usepackage{bm}
\usepackage{amsmath}
\usepackage{amssymb}
\usepackage{amsthm}
\newcommand{\be}{\begin{equation}}
\newcommand{\ee}{\end{equation}}
\newcommand{\bea}{\begin{eqnarray}}
\newcommand{\eea}{\end{eqnarray}}

\newcommand{\ph}[1]{\phantom{#1}}

\newcommand{\con}{\circ}

\begin{document}

\title{Inflation and dark energy from three-forms}
%\title{Three-form cosmology: Background and perturbation dynamics}

\author{Tomi S. Koivisto}
\email[]{t.koivisto@thphys.uni-heidelberg.de}
\affiliation{Institute f\"ur Theoretische Physik, Universi\"at Heidelberg, Philosophenweg 16, 69120 Heidelberg, Deutschland}
\author{Nelson J. Nunes}
\email[]{n.nunes@thphys.uni-heidelberg.de}
\affiliation{Institute f\"ur Theoretische Physik, Universi\"at Heidelberg, Philosophenweg 16, 69120 Heidelberg, Deutschland}

\date{\today}

\begin{abstract}

Three-forms can give rise to viable cosmological scenarios of inflation and dark energy with potentially observable signatures distinct from standard single scalar 
field models.
In this study, the background dynamics and linear perturbations of self-interacting three-form cosmology are investigated. 
The phase space of cosmological solutions possesses (super)-inflating attractors and saddle points which 
can describe three-form driven inflation or dark energy. The quantum generation and the classical evolution of 
perturbations is considered. The scalar and tensor spectra from a three-form inflation and the impact from 
the presence of a three-form on matter perturbations are computed. Stability properties and equivalence of the model with 
alternative formulations are discussed.

\end{abstract}


\maketitle

\section{Introduction}
\label{introduction}

Inflation is a successful explanation of many cosmological puzzles, and the current acceleration of the universe is a cosmological 
puzzle which yet lacks an 
explanation. Since Nordstr\"om \cite{Nordstrom:1988fi}, scalar fields have been present in extra dimensional and fundamental theories, and it is natural
to employ them to describe the energy sources needed to generate inflation and dark energy \cite{Brans:1961sx,Starobinsky:1980te,Wetterich:1987fm,
Peebles:1987ek,Amendola:1999er,Koivisto:2006xf}, for recent reviews, see \cite{Bassett:2005xm,Copeland:2006wr}.
However, it is crucial to understand how strict are the theoretical and phenomenological limits on the role of higher spin fields in cosmology. 

Vector inflation \cite{Ford:1989me} has been considered recently, using either time-like \cite{Koivisto:2008xf} or space-like \cite{Golovnev:2008cf} 
components. However, to naturally inflate, the vector needs a nonminimal coupling and seems to feature instabilities \cite{Himmetoglu:2008zp}, see 
however \cite{Karciauskas:2008bc,Golovnev:2009ks}. Effects on CMB, alternative scenarios \cite{Watanabe:2009ct} and the perturbation generation 
\cite{Yokoyama:2008xw,Dimopoulos:2008yv,Dimopoulos:2009am} have been studied in these 
models. Vector field dark energy \cite{Kiselev:2004py,ArmendarizPicon:2004pm,Koivisto:2005mm,Boehmer:2007qa,Koivisto:2007bp} might alleviate the  
coincidence problem \cite{Wei:2006tn,Koivisto:2008ig,Jimenez:2009ai} and introduce new effects on perturbations \cite{Mota:2007sz,Dulaney:2008ph}. It can be shown 
that two-form 
inflation resembles much the vector inflation, having the same 
possibilities and problems \cite{Germani:2009iq,Kobayashi:2009hj,Koivisto:2009sd}. Spinor \cite{Saha:2006iu,Boehmer:2007ut} and Yang-Mills 
\cite{Zhao:2005bu,Bamba:2008xa} fields have been also explored. Kalb-Ramond forms with dilaton 
couplings has been considered in the frameworks of 
string cosmology dynamics \cite{Copeland:1994km,Lukas:1996iq}, pre-big bang cosmology \cite{Gasperini:1998bm}, 
unified models of dark matter and dark energy \cite{Bilic:2005zk,Bilic:2006cp} and bouncing cosmology \cite{DeRisi:2007dn}.
Two-forms appear also in the asymmetric gravity \cite{Moffat:1994hv,Damour:1992bt} as the antisymmetric contribution to the metric, and have been considered in 
cosmology \cite{Prokopec:2005fb,Janssen:2006nn}. 
Recently a Chern-Simons type gravity was developed promoting the Levi-Civita symbol into a dynamical field \cite{Gupta:2009jy}. 
In the two-measures field theory, the new measure of integration can be built from either four scalar fields or an
independent dynamical three-form \cite{Guendelman:2008ms}. This theory has several cosmological implications \cite{Guendelman:2006af}. In a scale-invariant 
realization of the theory, a scalar potential acquires two flat regions in such a way that both inflation and dark energy may 
emerge \cite{Guendelman:1999qt,Guendelman:2002js}. Forms, being intrinsically anisotropic, could also be relevant for a dynamical origin of the four large 
dimensions \cite{ArmendarizPicon:2003qw,Gunther:2003zn,Saidov:2006xr}, modeling violation of the Lorenz invariance 
\cite{Jacobson:2000xp,Kostelecky:2000mm,Li:2007vz,Zuntz:2008zz,Carroll:2009en,Campanelli:2009tk}, 
the observed CMB anomalies \cite{OliveiraCosta:2003pu,Schwarz:2004gk,Eriksen:2004iu,Samal:2007nw,Hoftuft:2009rq} or testable late-time anisotropic 
phenomenology \cite{Blomqvist:2008ud,Cooray:2008qn,Koivisto:2007sq,Quercellini:2008ty,Quercellini:2009ni}.

In the present paper, our aim is to study the possible cosmological significance of three-forms. 
It was noticed in \cite{Duff:1980qv} that the four-form constructed from a three-form gauge potential generates a cosmological constant. Since then, this 
fact has been employed in discussions attempting to explain the tiny (or vanishing) value of the cosmological constant \cite{Hawking:1984hk,Turok:1998he}. 
Recently, we have proposed to consider the case of self-interacting gauge potential \cite{Koivisto:2009sd,Koivisto:2009ew}. This breaks the 
gauge-invariance but the field becomes then dynamical. Then a single field inflation with an exit to radiation 
dominated era can be naturally generated, or alternatively, the three-form can act as possibly transient dark energy at a late stage of the history of the universe
(three-from induced potentials were discussed in \cite{Gubser:2000vg,Frey:2002qc}). 
Form fields appearing in string theory generically couple to branes and this way a potential term might be obtained. 
In the present study we confine our investigations to the simple model with only a canonical field minimally 
coupled to Einstein gravity. A three-form generalization of vector (and scalar) inflation was introduced recently \cite{Germani:2009iq}, based on an action involving nonminimal couplings in such a way that the equation of motion of the comoving field has exactly the Klein-Gordon form in FLRW spacetime. 
The study of gravitational waves in such model reveals an instability occurring at large values of the field \cite{Kobayashi:2009hj} while the 
spectrum of scalar perturbations in small field inflation could be slightly red tilted and thus compatible with observations \cite{Germani:2009gg}. 
In the minimally coupled model that we consider here the equation of motion can be also written in the Klein-Gordon form, but given an effective potential. 

The dual of the three-form is a scalar field. In the case of a non-quadratic potential, the kinetic term of the scalar field is noncanonical.
Such a model then becomes equivalent to k-inflation, and has been analyzed before \cite{ArmendarizPicon:1999rj,Gruzinov:2004rq}. A non-quadratic
dependence on the three-form Faraday term results in a self-coupling of the scalar field. For any non-minimal coupling, in particular 
nonminimal gravity couplings of the three-form, the duality with a scalar field  breaks down. In fact, as we will show explicitly, 
even for some fairly simple self-interactions of the three-form, no dual description can be established in terms of a scalar field.
In any case, typically the scalar field description, if it exists, is opaque and intractable even if the model in three-form language is 
simple and intuitive. In fact most of the models admit a reformulation as vector models, and in some cases a presentation as 
 dynamical four-form models is also possible. Thus the simple starting point we have opens new perspectives on several classes of form 
cosmologies. 

We will write down the basic equations in section \ref{Applications} and review some results of Ref.~\cite{Koivisto:2009ew} to give an intuitive picture of the 
possible background dynamics. A convenient variable to describe is the comoving field $X$.
An analogy between a scalar field and the comoving field $X$ can be utilized to illustrate the behavior of the field, however, since the kinetic term of 
the actual field $X/a^3$ governs some aspects of the dynamics, an alternative viewpoint is also necessary to fully understand the dynamics. 
In section \ref{phase} we give a detailed account of the background 
expansion dynamics. The phase space analysis of the system reveals three classes of fixed 
points, one corresponding to matter domination and two corresponding to the domination of the form field. The nature and stability of the latter two 
points, which are relevant for dark energy and inflation solutions, depends on the form of the potential. We consider exponentials, power-law and 
Ginzburg-Landau type potentials. 

In section \ref{perturbations} we consider perturbations of three-form cosmology. Guided by the duality with a vector field, we parameterize the 
four degrees of freedom in the fluctuations of the three-form, two of them transforming as vectors under spatial rotations and two as scalars. It turns out that 
the vector-type fluctuations can be neglected about a FLRW background, whereas the scalar perturbations introduce several possible new effects in cosmology. We 
analyze the quantum generation of three-form perturbations during inflation using the standard techniques, and give the detailed general form of 
predictions for the
amplitude of scalar and tensor fluctuation spectrum and their spectral indices. The possible influence on matter inhomogeneities due to the presence of classical 
perturbations in a three-form in a matter-dominated universe is also considered. It is found that depending on the sound speed of the $X$-component (which in turn 
depends on the chosen form of the potential), there can be a range of scales where the linear growth of matter density perturbation is affected by the three-form 
fluctuation. This effect can be encoded into an effective strength of gravitational coupling of matter particles, 
which in general depends on time and length scale. 

Finally we discuss some formal aspects of the model.
In section \ref{Formalities}, some manipulations of the action are performed in order to clarify the properties of the 
model. The degrees of freedom and their nature will be seen to depend strongly on the form of the self-interactions. For some specific cases, 
dualities and equivalences can be established with other form field models as stated above. The model is also discussed briefly in the wider context
of general (quadratic) three-form actions. 
We conclude in section \ref{conclusions} stating a few central formulas we have obtained. Some details of the scalar field formulation in particular cases is given in 
the Appendix \ref{ap_s}, 
and an alternative viewpoint employing the dual vector is mentioned in appendix \ref{ap_v}.

%%%%%%%%%%%%%%%%%%%%%%%%%%%%%%%%%%%%%%%%%%%%%%%%%%%%%%%%%%%%%%%%%%%%%%%%
\section{Expansion dynamics}
%%%%%%%%%%%%%%%%%%%%%%%%%%%%%%%%%%%%%%%%%%%%%%%%%%%%%%%%%%%%%%%%%%%%%%%%
\label{Applications}

We shall focus on a canonical theory minimally coupled to Einstein gravity. The action for a three-form $A$ can then be written
\be \label{action}
S_A = -\int d^4 x \, \sqrt{-g}\left(\frac{1}{2\kappa^2}R - \frac{1}{48}F^2 - V(A^2)\right) \,,
\ee
where $\kappa^2 = 8\pi G$. In this section, to avoid unnecessary and excessive use of indices \footnote{See figure 3.1 of Ref. 
\cite{Eguchi:1980jx}.}, we introduce the following notations: squaring means contracting the indices in order, as $A^2 = 
A_{\alpha\beta\gamma}A^{\alpha\beta\gamma}$, 
dotting means contracting the first index, as $(\nabla\cdot A)_{\alpha\beta} = \nabla^\mu A_{\mu\alpha\beta}$, and circling means 
contracting all but the first index in order, $(A \con B)_{\mu\nu} = A_{\mu\alpha\beta}B_\nu^{\ph{\nu}\alpha\beta}$ and finally 
antisymmetrization is performed by square brackets, for example $[A]_{\mu\nu} = \frac{1}{2}(A_{\mu\nu}-A_{\nu\mu})$. Now the possible 
drawback seems to be that the valence of the objects we are dealing with is not explicitly seen; however, in this index-free notation most 
results automatically generalize for tensors of arbitrary valence. In this notation the $F(A)$ is $F=(n+1)[\nabla A]$, where $A$ is a 
$n$-form, and thus $F(A)$ generalizes the Faraday form appearing in Maxwell theory.
The energy momentum tensor is
\be \label{emt}
T = \frac{1}{6}F\con F + 6V(A^2)A\con A - g\left(\frac{1}{48}F^2+V(A^2)\right). 
\ee
The action leads to the equations of motion
\be \label{eom}
\nabla \cdot F = 12V'(A^2)A,
\ee
which implies, due to antisymmetry, the additional set of constraints
\be \label{const}
\nabla \cdot V'(A^2)A=0.
\ee

We consider a flat FLRW cosmology described by the line element
\be
ds^2 = -dt^2+a^2(t)d{\bf x}^2.
\ee 
The nonzero components of most general three-form compatible with this geometry are then given by
\be
A_{ijk}  =  a^3(t)\epsilon_{ijk}X(t). 
\ee
where we have considered, instead of the field $A$, the more convenient comoving field $X$, and $i$, $j$, $k$ denote the spatial indices. The relation between the 
squared invariant $A^2$ and the 
comoving field $X$ is then $A^2 = 6X^2$. We can thus consider the potential as a function of $X^2$, as we will do 
in the following. 

The equation of motion of the field $X$ is then:
\be
\label{eomX}
\ddot{X} = - 3 H \dot{X} - V_{,X} - 3 \dot{H} X  \,.
\ee
An overdot means derivative with respect to the cosmic time $t$.
The background perfect fluid evolves with
\be
\dot{\rho}_B = -3\gamma H\rho_B \,,
\ee
where $\gamma = 1 + p_B/\rho_B$,   
and these equations are subject to the Friedmann constraint
\begin{eqnarray}
H^2 &\equiv& \left(\frac{\dot{a}}{a}\right)^2 \nonumber \\
     &=& \frac{\kappa^2}{3} \left( \frac{1}{2} (\dot{X}+3H X)^2 + V(X)+ \rho_B \right) \,. \label{f1} 
\end{eqnarray}
The other field equation follows also from this and the continuity equations as
\be
\dot{H} = -\frac{\kappa^2}{2} \left(V_{,X}X + \gamma \rho_B\right) \label{f2} \,.
\ee
We can thus define energy density and pressure of the field as
\begin{eqnarray}
\rho_X &=& \frac{1}{2} (\dot{X}+3H X)^2 + V(X) \,, \\
p_X &=& -\frac{1}{2} (\dot{X}+3H X)^2 - V(X) + V_{,X} X \,. \\
\end{eqnarray}
The equation of state parameter of the three-form, $w_X = p_X/\rho_X$,  can be written as
\be
\label{eos}
w_X = -1 + \frac{V_{,X}X}{\rho_X} \,.
\ee
We thus see directly that whenever the potential or just its slope vanishes, the field is like a cosmological constant. 
Furthermore, whenever the slope of $V(X)$ is negative (positive) if $X$ is positive (negative), the comoving field behaves as a phantom field. So 
the origin has some absolute meaning for this field, unlike in the case of a scalar. We also note that the equation of state is unbounded from
both up and below. 

\begin{figure}
\includegraphics[width=8.5cm]{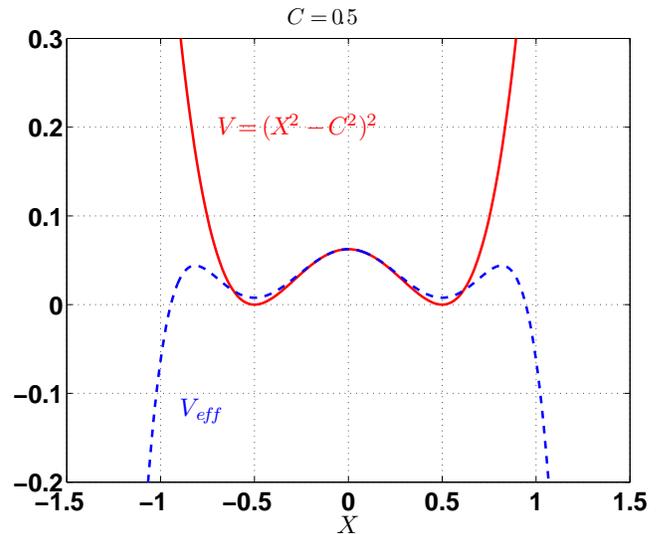}
\caption{\label{veff2} 
The potential (red, solid line) and the effective potential (blue, dashed line) for the potential $V=(X^2-C^2)^2$, when 
$C=\frac{1}{2}<\sqrt{\frac{2}{3}}$ and $\rho_B = 0$. Units of $\kappa = 1$.}
\end{figure}

\begin{figure}
\includegraphics[width=8.5cm]{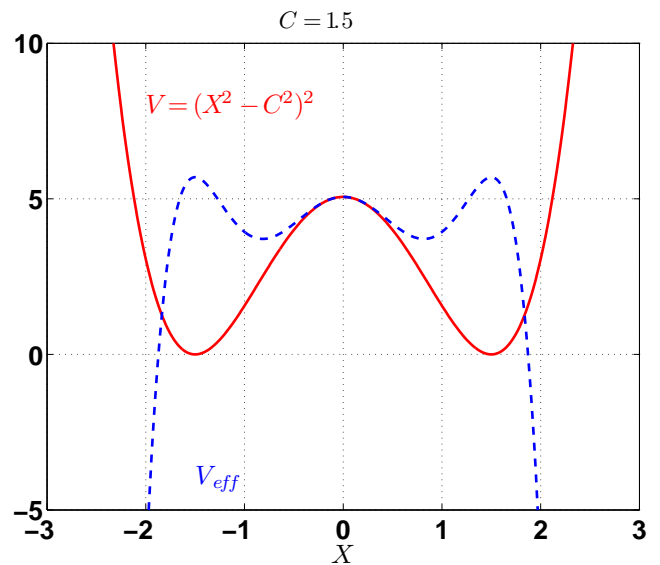}
\caption{\label{veff3} 
The potential (red, solid line) and the effective potential (blue, dashed line) for the potential $V=(X^2-C^2)^2$, when 
$C=\frac{3}{2}>\sqrt{\frac{2}{3}}$ and $\rho_B = 0$. Units of $\kappa = 1$.
}
\end{figure}

Previously it was shown that one may predict the evolution of the system by considering the effective potential, defined by \cite{Koivisto:2009ew}
\begin{eqnarray} \label{v_eff}
{V_{\rm eff}}_{,X} = V_{,X} + 3 \dot{H} X.
\end{eqnarray}
We illustrate the form of this potential in Figs. \ref{veff2} and \ref{veff3} for two new cases when $\rho_B = 0$.
For these potentials, the positions of the minima depend on the precise value of $C$. When $C < \sqrt{2/3}$, the local minima are 
at $X = \pm C$ and when $C > \sqrt{2/3}$ the minima are at $C = \pm \sqrt{2/3}$. 
This suggests that the field has different dynamics given a choice of $C$. In particular it seems that the late time value of $X$ is $\pm C$ if $C < \sqrt{2/3}$ and 
$X$ approaches $\sqrt{2/3}$ if $C > \sqrt{2/3}$. We will see in the Section \ref{phase} that these first impressions are indeed correct.
One notes that there are places where the slope of the effective potential is downwards while the bare potential is increasing. In such situations the field,
as it rolls down the effective potential, climbs up its bare potential. The analogy with the scalar field holds here in the sense that such cases indeed correspond 
to a phantom-like expansion, where the (effective) equation of state of the field is more negative than minus unity. 

\begin{figure}
\includegraphics[width=8.5cm]{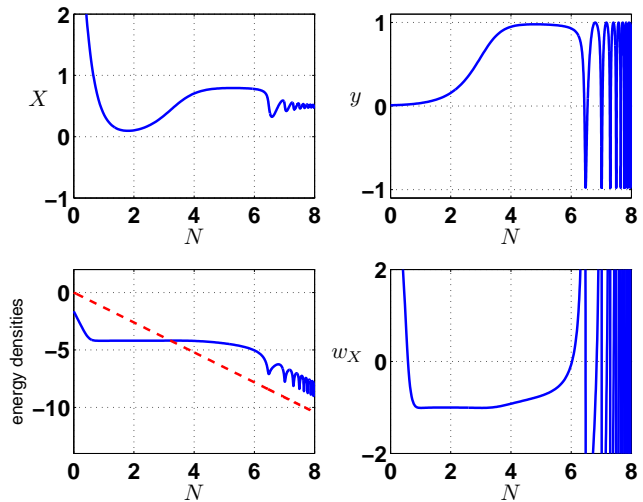}
\caption{\label{x1} 
Cosmological evolution as a function of the e-folding time $N=\ln{a}$ for the Landau-Ginzburg potential when $C=0.5$ (units of $\kappa = 1$). In the upper left 
panel we see the field going through the unstable fixed point at $X=\sqrt{2/3}$ to oscillate around the stable minimum at $X=C$. The upper right panel shows 
the corresponding values of the potential. The lower left panel depicts the energy densities in a logarithmic scale; the dashed (red) line is matter and the 
solid (blue) line is the three-form. A brief tracking period is included in the figure, followed by two stages: the unstable fixed point and the 
oscillations around the stable minimum. The lower right panel shows the equation of state of $X$, exhibiting strong oscillations during the settling to the minimum. }
\end{figure}

\begin{figure}
\includegraphics[width=8.5cm]{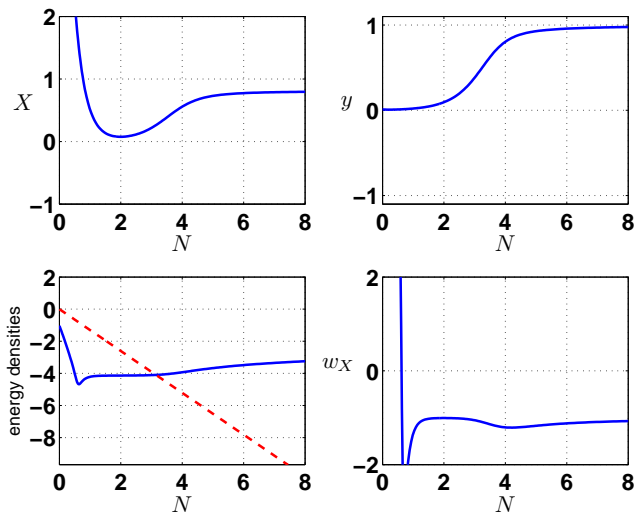}
\caption{\label{x2} 
Cosmological evolution as a function of the e-folding time $N=\ln{a}$ for the Landau-Ginzburg potential when $C=1.5$ (units of $\kappa = 1$). In the upper left 
panel we see the field going through the unstable fixed point at $X=0$ to settle to the stable minimum at $X=\sqrt{2/3}$. The upper right panel shows 
the corresponding values of the potential. The lower left panel depicts the energy densities in a logarithmic scale; the dashed (red) line is matter and the 
solid (blue) line is the three-form. A brief tracking period is included in the figure, followed by two stages: the unstable fixed point and the 
approach to the late time attractor $X = \sqrt{2/3}$. The lower right panel shows the equation of state of $X$, exhibiting a phantom behavior during the approach to the stable 
minimum. }
\end{figure}

We illustrate these behaviors also with numerical solutions.
In Figs.~(\ref{x1}) and (\ref{x2}), we show the evolution of the field, energy density and equation of state for the Landau-Ginzburg potential with $C = 0.5$ and $C = 1.5$, respectively. 
We use this potential because it illustrates most features of the dynamics. We will see that depending on the magnitude of $C$, two qualitatively different cases 
emerge. Though the early evolutions are similar while the background is dominant and the potential is roughly $V \approx X^4$, the late time evolutions diverge. 
This is to be expected because, as we have seen in Figs.~(\ref{veff2}) and (\ref{veff3}), the effective potentials are different. In particular, we see that the 
field settles in $X = C$ in the first case as $C < \sqrt{2/3}$ and the late time oscillations result in the equation of state parameter to cross $w_X = -1$ at each oscillation. This can be understood by noticing that the field is transiting between positive and negative values of the slope of the potential at the minimum. For the second case $C > \sqrt{2/3}$, the field cannot reach its minimum as it is constrained to $X < \sqrt{2/3}$ for positive velocity. 
In particular, we see that $X$ cannot be displaced further than $X = \pm\sqrt{2/3}$ for $\dot{X} = 0$, as this saturates the Friedmann equation (\ref{f1}) and larger
field values could be reachable only in the presence of negative energy sources.
Consequently the equation of state approaches asymptotically $w = -1$ from below in the second case where $C$ is larger than the critical value.

In Figs.~\ref{x1} and \ref{x2} we see that at early times the evolutions are 
identical and initially 
are described by a tracking behavior followed by a constant energy 
density of the field $X$ and only at late times the evolutions diverge from one another. The earlier history of these evolutions corresponds to power-law 
potential regime of the potential, which was discussed in detail in Ref. \cite{Koivisto:2009ew}, but we make here brief comments from the dual point of view. 

It was shown that when the background fluid is dominant, and the one independent component of the three-form $A_{ijk}$ is constant,
the field will be tracking with $n = \gamma$, where $n$ is given by the shape of the potential, $V(X) = V_0 X^n$.
This period is ended when $3 H^2 y^2 \approx V(X)$. It was found that the number of e-folds elapsed is then rather accurately given by 
\be
N_s = \frac{1}{3n} \ln \left( \frac{V_i}{3 H^2_i y_i^2} \right) \,.
\ee
where now $y_i$ is the initial value of $y \equiv \kappa(X'+3X)/\sqrt{6}$ and $V_i/H_i^2$ is the initial ratio of the potential and the Hubble rate and prime means differentiation with respect to 
\be
N \equiv \ln a \,.
\ee
The numerical solutions shown here also follow this behavior now corresponding to the case $n=4$.

One may consider the tracking property in light of the dual description as a scalar field which we discuss in more generality in section 
\ref{duality}.  
Now such a dual description as a scalar field exists, since during the tracking phase the kinetic term is negligible and the potential can be approximated 
by a power-law. Then this dominating potential term represents the kinetic term of the scalar field. It can be easily shown that a k-essence Lagrangian $\mathcal{L} 
\sim (\partial\phi)^{2p}$ behaves as a perfect fluid with the equation of state given by $w_\phi = 1/(2p-1)$. During the tracking phase described above, the 
three-form indeed is (at least approximately) equivalent to such a k-essence field. One may deduce from the action (\ref{0form}) that a power-law potential
$V \sim A^n$ turns into the kinetic Lagrangian with $2p=n/(n-1)$, thus assuming the scaling of energy density we have obtained.

Canonical quintessence is also known to possess the tracking property in some cases. One might thus be curious if this fact could be used to construct tracking 
three-form models. It turns out to be the case, and to produce noncanonical three-forms models. Among the simplest examples of a quintessence with a tracking 
attractor is an inverse power-law potential $V(\phi)=V_0\phi^p$. It is known to approximately track the background density rather independently of the 
initial conditions. The power-law form of the potential translates into a power-law kinetic term of a three-form $\Psi$,
\be
\mathcal{L} = \left(\frac{x^p}{V_0 p^p}\right)^\frac{1}{p-1}(p-1)-\frac{1}{18}\Psi^2, 
\ee
where the kinetic term is given by
\be
x \equiv -\frac{1}{4}\epsilon_{\alpha\beta\gamma\delta}F^{\alpha\beta\gamma\delta}(\Psi)
\ee
The exponential potential is known to be the special case possessing the scaling, or ''exact tracking'' property. In section \ref{duality} we will describe how 
the three-form can sometimes be written as quintessence; by going the other way around one may find that a quintessence model specified by minimal coupling and $V(\phi) = V_0 e^{-\lambda \phi}$ 
can be recast into the three-form model
\be
\mathcal{L} = -\frac{x}{\lambda}\left(1+\ln{\frac{x}{\lambda V_0}}\right) - \frac{1}{18}\Psi^2.
\ee
In the present study we however confine to the canonic kinetic term. As shown above, the three-form energy density can then scale as a power of the scale 
factor, given the initial condition that the (only independent) component of the field, $A_{ijk}$ is a constant, meaning that the comoving field scales 
as $X\sim a^{-3}$. 

Before settling into the minimum, the field turns around the potential and start climbing it. The value of $N_t$ when this happens is given by
\be
N_t = \frac{1}{3\left(1+\gamma/2\right)} \, \ln \left(\frac{2}{\gamma} \frac{B}{A}\right) \,,
\ee
where $A=\sqrt{2/3} y_i/(1+\gamma/2)$, and $B=X_i-A$, where $X_i$ is the initial value of the field. To have scaling behaviour for many e-folds, one may consider tiny $A$ at huge $X_i$. 

Let us then consider the transient acceleration scenario.
There the field slows down never reaching the critical value $X=\sqrt{2/3}$, the amount of inflation $N_e$ is eventually given by
the initial value of the field $X_i$ at the beginning of inflation near the critical point and again depends on the slope shape,
\be
\kappa X_i = \pm \left(\frac{2}{3} - \frac{4}{9n} \, \frac{1}{1+2 N_e}\right)^{1/2} \,,
\ee
and clearly, the slow-roll condition on the velocity of the field (\ref{dx1}) must be satisfied. We observe that for larger values of $N_e$ then $|\kappa X_i|$ must be  closer to $|\kappa X_i| \approx \sqrt{2/3}$. 

We also showed in Ref.~\cite{Koivisto:2009ew} that the oscillations of the field, when it settles to the minimum, follow an averaged behavior which depends on the 
shape of the potential. For a power law potential the result is, surprisingly, the same as for a canonical scalar field \cite{1983PhRvD..28.1243T},
\be
\langle w_X \rangle = \frac{n-2}{n+2} \,.
\ee
Thus for $n = 2$ the fields behaves as dust, $\langle w_X \rangle = 0$ and for $n = 4$ it mimics radiation, $\langle w_X \rangle = 1/3$ \footnote{The nontrivial 
case of scalar potential including both quartic and quadratic terms has been solved just recently.}. We mention in passing that this
result applies also for oscillating k-essence with mass term and a power-law kinetic term due to the duality mentioned above, so the model 
$\mathcal{L} \sim (\partial\phi)^{2p} + m^2\phi^2$ oscillates like $\langle w_\phi \rangle = (p-1)/(3p-1)$.

This concludes our qualitative review of the possible sequences of cosmological epochs, and next we turn into more rigorous phase space analysis
and observable predictions.   

%%%%%%%%%%%%%%%%%%%%%%%%%%%%%%%%%%%%%%%%%%%%%%%%%%%%%%%%%%%%%%%%%%%%%%%%%%
\section{Phase space analysis}
%%%%%%%%%%%%%%%%%%%%%%%%%%%%%%%%%%%%%%%%%%%%%%%%%%%%%%%%%%%%%%%%%%%%%%%%%%

\label{phase}

In this section we will put on a more solid and formal ground the considerations of the previous section on the late time dynamics of the system and its stability.
We start by rewriting the equations of motion in the form of a system of first order differential equations
\begin{eqnarray}
\label{xeq}
x' &=&  3 \left(\sqrt{\frac{2}{3}}y - x \right) \,, \\
\label{yeq}
y' &=& -\frac{3}{2} \lambda(x) \left(1-y^2-w^2\right) \left[x y - \sqrt{\frac{2}{3}} \right] + \frac{3}{2} \gamma w^2 y \,, \\
\label{weq}
w' &=& -\frac{3}{2} w \left(\gamma + \lambda(x) \left(1-y^2-w^2\right) x - \gamma w^2 \right) \,,
\end{eqnarray}
%
%where a prime means differentiation with respect to $N \equiv \ln a$, where $a$ is the scale factor of the universe and 
where we have defined
\begin{eqnarray}
x \equiv \kappa X \,, \hspace{0.7cm} y \equiv \frac{\kappa}{\sqrt{6}} (X'+ 3X) \,, 
\hspace{0.7cm} z^2 = \frac{\kappa^2 V}{3H^2} \,,
\end{eqnarray}
\begin{eqnarray}
w^2 \equiv \frac{\kappa^2 \rho_B}{3 H^2} \,, \hspace{1cm} \lambda(x) \equiv - \frac{1}{\kappa} \frac{V_{,X}}{V} \,.
\end{eqnarray}
$\lambda(x)$ is, therefore, a function of $X$. 
The quantity $z$ was eliminated from the equations of motion by applying the Friedmann constraint
\be
\label{friedxy}
y^2 + z^2 + w^2 = 1 \,.
\ee

The system (\ref{xeq})--(\ref{weq}) has three critical points which are described in Table~\ref{tb1}.

\begin{table}
\begin{tabular}{|c|c|c|c|c|c|c|c|}
\hline
~ & $x$ & $y$ & $w$ & $\dot{H}/H^2$ &$\lambda$  & description \\ \hline 
A & 0 & 0 & $\pm 1$ & $-3\gamma/2$ &any&  matter domination \\ \hline
${\rm B}_{\pm}$ & $\pm \sqrt{2/3}$ & $\pm1$ & 0 & 0 & any &  maximal point  \\ \hline
C & $x_{\rm ext}$ & $\sqrt{3/2}\, x_{\rm ext}$ & 0 & 0 & $0$ & potential extremum \\ \hline
\end{tabular}
\caption{\label{tb1}} The critical points in the system.
\end{table}

{\bf A}: $x = 0$, $y = 0$, $w = \pm1$, for any $\lambda$. It corresponds to the background dominated solution. At late time the ratio $\dot{H}/H^2$ approaches $-3\gamma/2$. If $\gamma$ is a constant, the eigenvalues are 
$(-3,3\gamma/2,3\gamma)$, hence, it is an unstable critical point. 

{\bf B}: $x = \pm \sqrt{2/3}$, $y = \pm1$, $w = 0$, for any $\lambda$. This is a critical point that does not exists for the standard scalar field and that results from the extra $X$ dependent terms in the equation of motion and in the definition of the energy density and pressure. When approaching this fixed point, 
$\dot{H}$ approaches a constant at late times, however, $H^2$ keeps increasing, therefore, the effective equation of state parameter of the field $X$ 
approaches $-1$ from below. The eigenvalues are $(-3,0,-3\gamma/2)$ and because one of the eigenvalues is zero, we cannot infer anything about the nature of the critical point from the 
linear analysis. We need to consider specific potentials and go to nonlinear order. The eigenvector corresponding to the vanishing eigenvalue reads $(\sqrt{2/3},1,0)$, therefore, when going to higher order we study the stability of perturbations along the zero eigenvalue direction $\delta r=\sqrt{2/3} \,\delta x + \delta y$, for which we get
\be
\delta r' = \mu^{(n)} \, \delta r^n \,,
\ee
for $n > 1$ and $\mu^{(n)}$ is the coefficient resulting from expanding  equations Eqs.~(\ref{xeq}) and (\ref{yeq}) to $n$th order 
and using $\delta x = \sqrt{6}\,\delta r/5$ and $\delta y = 3 \,\delta r /5$, such that $\mu^{(1)} =1$.
The general solution to this equation is
\be
\delta r = \delta r_0 \,\left(1- \delta r_0^{n-1} \, (n-1) \, \mu^{(n)} \, N \right)^{1/(1-n)} \,.
\ee
In order for a negative initial perturbation ($\delta r_0 <0$) to decay one must have $\mu^{(n)} > 0$ if $n$ is even and $\mu^{(n)} < 0$ if $n$ is odd. For a positive perturbation it suffices to have $\mu^{(n)} < 0$, regardless of the value of $n$.

Since $|y|$ must be less than unity, there can only be negative perturbations along the $r$ direction about the fixed point $(x,y,w) = (\sqrt{2/3},1,0)$ and positive perturbations about the fixed point $(x,y,w) = (-\sqrt{2/3},-1,0)$, thus, for the perturbations to decay it is required that $\mu^{(n)}$ must be positive for fixed point $(x,y,w) = (\sqrt{2/3},1,0)$ if $n$ is even and $\mu^{(n)}$ negative if $n$ is odd and $\mu^{(n)}$ negative for fixed point $(x,y,w) = (-\sqrt{2/3},-1,0)$.

{\bf C}: $x = x_{\rm ext}$, $y = \sqrt{3/2} \, x_{\rm ext}$, $w = 0$ where $x_{\rm ext}$ corresponds to the value of $x$ at the extrema of the potential, i.e. where 
$\lambda = 0$. In this case, $H^2$ becomes constant and $\dot{H}$ vanishes at late times. The stability of these fixed points, is therefore, strongly dependent on the specific form of the potential.

We shall now look at particular examples to illustrate the significance and stability of the fixed points just described.

\subsubsection{$V = \exp(-\beta X)$}

Because the potential, being a function of the invariant $A^2$, must depend explicitly on $X^2$ instead of $X$ itself, we are dealing with symmetric 
potentials. This potential is not $X^2$ dependent and therefore should only be seen as an example to be compared with the standard scalar field dynamics.
We can compute $\mu$ for both fixed points ${\rm B}$ to find that 
\be
\mu^{(2)}_{{\rm B}_\pm} = \frac{18}{25} \sqrt{6} \beta \,,
\ee
hence, $(\sqrt{2/3},1,0)$ is stable for $\beta > 0$ and $(-\sqrt{2/3},-1,0)$ is stable for $\beta < 0$. 

The fixed point C which corresponds to vanishing derivative of the potential.
Since we have assumed that $\beta=\lambda$ is a constant, in this case the potential is flat ($\lambda=0$) in some finite region of X, instead of a point we 
actually have a curve $C$. Then its points live in an effectively two dimensional manifold, and one of the Lyaponov exponents is expected to vanish.
Therefore we can now infer the nature of the critical point from the linear analysis. Since the two nonzero eigenvalues are negative, we have what is
called a local sink. Note that in fact $\lambda=0$ corresponds to the massless field, and the reduction of the dimension of the phase space reflects the
disappearance of a degree of freedom in the massless case due to the restored gauge invariance.

\subsubsection{$V = \exp(-\beta X^2)$}
This potential has a $X^2$ dependence hence is of the type we are looking for. If $\beta$ is positive, it presents a maximum at $X = 0$ and conversely, if $\beta$ is
negative, it has a global minimum at the same value.

Now the value of $\mu$ for fixed points B are
\be
\mu^{(2)}_{{\rm B}_\pm}= \pm \frac{72}{25} \beta \,,
\ee
hence they are both stable if $\beta > 0$.

The fixed point C is in this case $(x,y,w) = (0,0,0)$ and its corresponding eigenvalues are $m_{1,2} = -(3/2)(1\pm \sqrt{1+8\beta/3})$ and $m_3 = -3\gamma/2$. This point is stable provided 
$\beta < 0$, which clearly makes sense as it is the case that leads to a minimum in the potential.

\subsubsection{$V = X^2 + k$} 
Here $k$ is a positive constant introduced for the purpose of regularization purposes about $x=0$. The quadratic potential is in a sense similar 
to the previous one with $\beta<0$. Indeed we find that for fixed points 
B,
\be
\mu^{(2)}_{{\rm B}_\pm} = \mp \frac{72}{24} \,\frac{1}{2/3 + k} \,,
\ee
which mean that they are unstable.
In this case, the fixed point C, also corresponds to $(x,y,w) = (0,0,0)$ with eigenvalues 
\begin{eqnarray}
m_{1,2} &=& -\frac{3}{2} \left(1 \mp \sqrt{1-\frac{8}{3k}}\right) \,, \\
m_3 &=& -\frac{3 \gamma}{2} \,,
\end{eqnarray}
thus this is a stable fixed point.

\subsubsection{$V = X^4 + k$}

Though the quartic potential seems very similar to the quadratic potential there are in fact some differences. Again $k$ is a positive constant. For fixed 
point B we compute
\be
\mu^{(2)}_{{\rm B}_\pm} = \mp \frac{96}{25} \, \frac{1}{4/9 + k} \,,
\ee
and again they are unstable.
Fixed point C, which is again at $(x,y,w) = (0,0,0)$ and also has eigenvalues $(-3,0,-3\gamma/2)$ like fixed point B where the direction of vanishing eigenvalue is still given by $\delta r = \sqrt{2/3} \, \delta x + \delta y$. Going to second order in perturbations along this direction, like we did for point B, we find that $\mu_{\rm C}^{(2)} = 0$. 
We thus have to go to the third order, 
\be
\mu^{(3)}_{{\rm C}} = - \frac{72}{125 \, k} \,,
\ee
and there we find that the sign of the eigenvalue is negative, thus, the point is a stable attractor.

\subsubsection{$V = \left(X^2-C^2\right)^2 + k$}

This potential has two minima at $X = \pm C$ and a maximum at $X = 0$ (we are taking $C > 0$). For fixed 
point B we can readily calculate
\be
\mu^{(2)}_{{\rm B}_\pm} = \mp \frac{144}{25} \, \frac{(2/3-C^2)}{(2/3-C^2)^2 + k} \,,
\ee
hence, both fixed points are stable provided $C > \sqrt{2/3}$.
We have now, however, three type C fixed points:
\begin{eqnarray}
{\rm C}_1: &~& \, \, \left(\pm C, \pm \sqrt{\frac{3}{2}} C,0 \right) \,, \\
{\rm C}_2: &~& \, \, (0,0,0) \,.
\end{eqnarray}
For ${\rm C}_1$ we find eigenvalues
\begin{eqnarray}
m_{1,2} &=& -\frac{3}{2} \left(1 \pm \sqrt{1-\frac{24 C^2}{k}\left(C^2 -\frac{2}{3}\right)^2} \right) \,, \\
m_3 &=& -\frac{3\gamma}{2} \,,
\end{eqnarray}
Thus, this fixed point is stable for $k > 24 C^2 (C^2-2/3)^2$. We must also point out that this fixed point only exists for $C < \sqrt{2/3}$ as we must require  $|y|<1$. 
For ${\rm C}_1$ we can compute the following eigenvalues:
\begin{eqnarray}
m_{1,2} &=& -\frac{3}{2} \left(1 \pm \sqrt{1+ \frac{16C^2}{3(C^4+k)}} \right) \,, \\
m_3 &=& -\frac{3\gamma}{2} \,,
\end{eqnarray}
consequently we can conclude that this fixed point is unstable which is not surprising given that it corresponds to the local maximum of the potential. 

The properties of the fixed points for all these forms of the potential are summarized in Table \ref{tb2}.

\begin{table*}
\begin{tabular}{|c|c|c|c|}
\hline
$V(X)$ & ~A~ & B & C \\ \hline
$\exp(-\beta X)$ & U & $B_{+}$ stable for $\beta > 0$ & S \\ 
~  & ~ & $B_{-}$ stable for $\beta < 0$ & ~ \\ \hline
$\exp(-\beta X^2)$ & U & stable for $\beta > 0$ & stable for $\beta < 0$ \\ \hline
$X^2$ & U & U & S \\ \hline
$X^4$ & U & U & S \\ \hline
$\left(X^2-C^2\right)^2$ & U & stable for $C > \sqrt{2/3}$ & $C_1$ stable, $C_2$ unstable \\ \hline 
\end{tabular}
\caption{\label{tb2} Stability of the fixed point in four classes of models. U -- unstable; S -- stable.}
\end{table*}

\begin{figure}
\includegraphics[width=8.5cm]{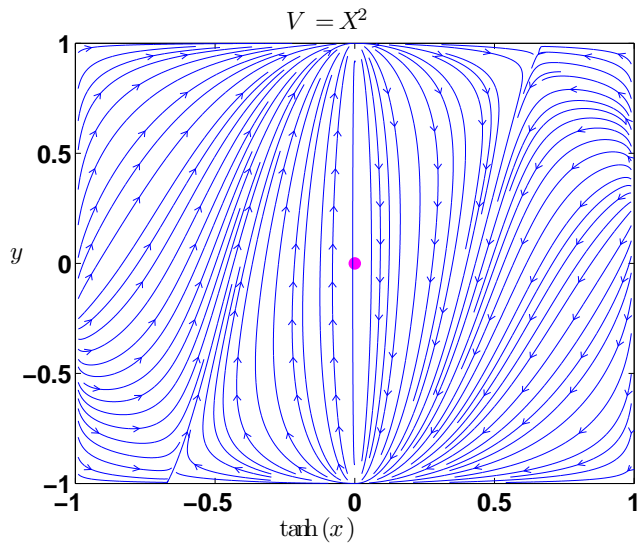}
\caption{\label{sp1} The phase space trajectories for quadratic potential.  The minimum of the potential at $X = 0$ has been marked with a dot.}
\end{figure}

\begin{figure}
\includegraphics[width=8.5cm]{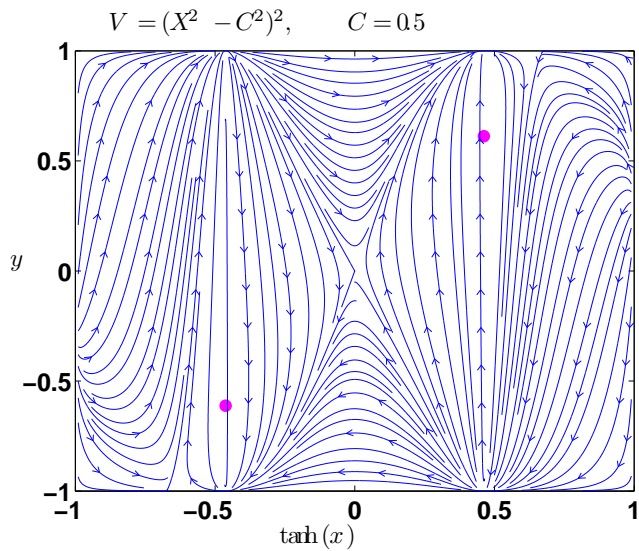}
\caption{\label{sp2} The phase space trajectories $V=(X^2-C^2)^2$, when $C=\frac{1}{2}<\sqrt{2/3}$. The stable fixed points at the minima of the potential
(both the true and the effective potential) have been marked with a dots.}
\end{figure}

\begin{figure}
\includegraphics[width=8.5cm]{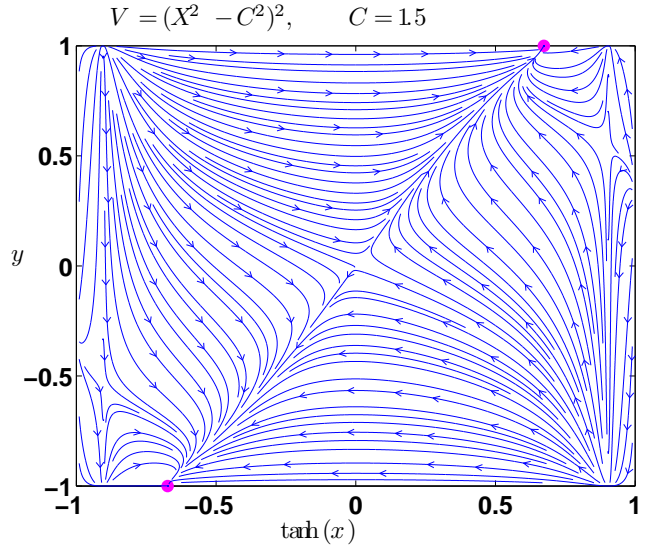}
\caption{\label{sp3} The phase space trajectories $V=(X^2-C^2)^2$, when $C=\frac{3}{2}>\sqrt{2/3}$. The stable fixed points at the minima of the effective
potential have been marked with a dot.}
\end{figure}

We are going to focus on two forms of the potential that will suffice to describe the general properties of this system that we have formally described above. 
In Fig.~\ref{sp1} we show the phase space portrait for a potential of the form $V(X) = X^2$. 
It shows that the trajectories approach the minimum of the potential at $X = 0$ and oscillate around this value. 
In Figs.~\ref{sp2} and \ref{sp3} we show the phase space trajectories for the potential $V(X) = (X^2-C^2)^2$, where $C$ is a constant. 
In the example of Fig.~\ref{sp2}, $C = 0.5$ and we see that the late time behavior consists of the trajectories oscillating around the fixed point 
$X = C$ and $y = 0$. Finally for the example of Fig.~\ref{sp3}, we see that the trajectories approach the critical points $x = \pm \sqrt{2/3}$, $y = \pm1$. 


%%%%%%%%%%%%%%%%%%%%%%%%%%%%%%%%%%%%%%%%%%%%%%%%%%%%%%%%%%%%%
\section{Cosmological perturbations}
\label{perturbations}

The general perturbations about the FLRW background can be parameterized by writing the line element as
\bea
ds^2 & = & -(1+2\psi)dt^2 + 2b_i dx^i dt \nonumber \\ & + & a^2(t)(1-2\phi)dx^i dx_i + a^2(t)h_{ij} x^ix^j,
\eea
where the two scalar perturbations $\psi$ and $\phi$ are the usual Bardeen potentials in the longitudinal gauge,
$b_i$ is a transverse vector and $h_{ij}$ is transverse and traceless as it describes the tensorial perturbations.

The field equations for the scalar perturbations are then
\be
-\frac{\nabla^2}{a^2}\phi + 3H(\dot{\phi} + H\psi)  =  -4\pi G \delta\rho, \label{e00} 
\ee
\be
-\frac{\nabla^2}{a^2}(\dot{\phi}+H\psi)  =  4\pi G (\rho+p)\frac{\theta}{a}, \label{e0i} 
\ee
\be
\ddot{\phi} + H(3\dot{\phi}+\dot{\psi}) + (2\dot{H}+3H^2)\psi - \frac{\nabla^2}{3a^2}(\phi-\psi) =  4\pi G \delta p, \label{eii} 
\ee
\be
-\nabla^2(\phi-\psi) =  12\pi G(\rho+p)\varpi. \label{eij}
\ee
The first equation is the energy constraint $(G^0_{\ph{0}0}$ component),
the second is the momentum constraint $(G^0_{\ph{0}i}$ component) involving the velocity perturbation $\theta$,
the third is the trace of the spatial part $(G^i_{\ph{i}i}$ component),
and the last one gives the shear propagation equations part $(G^i_{\ph{i}j}$ component) for the shear $\varpi$.

\subsection{The three-form}

To parameterize the fluctuations of the three-form, we employ a similar decomposition as for the metric. The four degrees of freedom in a three-form then 
turn
out to be two scalar and two vector degrees of freedom. The components of the three-form are fully specified by
\bea
A_{0ij} & = & a^2(t)\epsilon_{ijk}(\alpha_{,k} + \alpha_k) \label{0ij}, \\
A_{ijk} & = & a^3(t)\epsilon_{ijk}(X(t)+\alpha_0). \label{ijk} \\
\eea
Here $\alpha_k$ is a transverse vector and thus has two independent degrees of freedom. One easily sees that as usually, the vector and scalar
perturbations decouple at linear order. The square of the field is then
\be \label{square}
A^2 = 6\left[X^2+2X(\alpha_0+3X\phi)\right].
\ee
Under general gauge transformation $x^\mu \rightarrow x^\mu + \xi^\mu$, specified by the vector $\xi^\mu = (\xi^0,\xi_{,i}+\xi^i)$, where $\xi^i_{\ph{i},i}=0$, the 
field 
fluctuations transform as
\bea
\alpha_0 & \rightarrow & \alpha_0 - \dot{K}\xi^0 + X \nabla^2\xi, \label{g1} \\
\alpha & \rightarrow & \alpha - aX\dot{\xi}, \label{g2} \\
\alpha_i & \rightarrow & \alpha_i -aX\xi_i. \label{g3}
\eea
Here we use the variable $\dot{K} = \dot{X} + 3HX$.
In this transformation the metric potentials transform as
\bea
\psi & \rightarrow & \psi - \dot{\xi}^0, \\
\phi & \rightarrow & \phi + H\xi^0+\frac{1}{3}\nabla^2\xi.
\eea

The equations of motion for the scalar perturbations then are
\be
\dot{\alpha}_0 + 3H\alpha_0 + \frac{V_{,X}}{X}\alpha - \frac{\nabla^2}{a^2}\alpha + \dot{K}(3\phi-\psi) =  0, \label{a1} 
\ee
\bea
\ddot{\alpha}_0 & + & 3H\dot{\alpha}_0 + (3\dot{H} + V_{,XX})\alpha_0 - \frac{\nabla^2}{a^2}(\dot{\alpha}-2H\alpha) \label{a0} \\ \nonumber 
& + & \dot{K}(3\dot{\phi} - \dot{\psi}) + 3(V_{,XX}X-V_{,X})\phi + 2V_{,X}\psi  =  0. 
\eea
which represent the two independent components ($0ij$ and $ijk$) of Eq.~(\ref{eom}). 
The background equations of motion was used to simplify the second one. 
 As discussed in the previous section, the kinetic terms are given by this component, corresponding to the
field itself, while the potential depends on the comoving field $X$. This variable $\dot{K}$ is just the dimensional version of the $y$ we used in the phase 
space analysis. We notice also that the constraint equation (\ref{const}), which may be written as
\be \label{const2}
\frac{\partial}{\partial t}\left(\frac{V_{,X}\alpha}{X}\right) = V_{,X}(\alpha_0+3X\alpha)-V_{,X},
\ee
is not independent but follows consistently from Eqs.~(\ref{a1}) and (\ref{a0}).
The density, pressure and velocity perturbations come out as follows:
\bea 
\delta\rho_X  & =  & \dot{K}\left[\dot{\alpha}_0+3H\alpha_0 - \frac{\nabla^2}{a^2}\alpha + \dot{K}(3\phi-\psi)\right] \nonumber \\ 
& + & V_{,X}(\alpha_0+3X\phi), \label{dens} 
\eea
\be 
\delta p_X    =  -\delta\rho_X + (V_{,XX}X+V_{,X})(\alpha_0+3X\phi), \label{pres} 
\ee
\be
(\rho_X+p_X)\theta_X  =  -\frac{\nabla^2}{a^2}V_{,X}\alpha. \label{velo} 
\ee
\be
\varpi_X  =  0. \label{shear}
\ee
The first equation is the energy piece of the energy momentum tensor Eq.~(\ref{emt}) $(T^0_{\ph{0}0}$ component),
the second is the trace of the spatial part $(T^i_{\ph{i}i}$ component),
the third is obtained from the momentum containing part $(T^0_{\ph{0}i}$ component),
and the last one gives the symmetric traceless part of the spatial $T^i_{\ph{i}j}$ components.
Thus we get that the anisotropic stress due to the three-form vanishes.

In the absence of other vector sources, the rotational perturbations evolve like
\bea
\dot{b}_i + Hb_i & = & 0, \\
\frac{\nabla^2}{a^2}b_i & = & V_{,X}(Xb_i - \alpha_i).
\eea
Thus the vector perturbations decay and can be ignored. Since the three-form does not generate tensor perturbations, their evolution equation
is 
\be \label{grav}
\ddot{h}+3H\dot{h}-\frac{\nabla^2}{a^2}h = \varpi^{(t)},
\ee
where $\varpi^{(t)}$ is tensorial anisotropic stress to which the three-form does not contribute.

\subsection{Three-form domination}

Assume that the three-form dominates. Since it does not generate anisotropic stress, Eq.~(\ref{eij}) tells us that $\psi=\phi$. We can now derive an evolution 
equation
for the Bardeen potential $\phi$ in a closed form. Eq.~(\ref{0ij}) can be used to eliminate $\dot{\alpha}_0$ from the system.
Equations (\ref{e00}) and (\ref{dens}) may then be used to eliminate $\alpha_0$. Note that only $\alpha$ appears always without derivatives in these equations. We may
eliminate $\alpha$ using Eqs.~(\ref{e0i}) and (\ref{velo}). Finally, plugging the solutions into Eq.~(\ref{eii}) with the right hand side given by (\ref{pres}) we get
\begin{eqnarray} 
\label{phievol}
\ddot{\phi} + \left(H - \frac{\ddot{H}}{\dot{H}}\right)\dot{\phi}
+ \left(2\dot{H} - \frac{\ddot{H}H}{\dot{H}}\right)\phi  = \nonumber \\ \left(1-\frac{\ddot{H}}{\dot{H}}\frac{X}{\dot{X}}\right)\frac{1}{a^2} 
\nabla^2\phi  \,.
\end{eqnarray}
We can verify this result by differentiating Eqs. (\ref{e0i}) and (\ref{a1}) solving for $\dot{\alpha}$ and $\ddot{\alpha}_0$, and verifying that
the latter agrees with Eq.~(\ref{a0}). 
The RHS of Eq.~(\ref{phievol}) is simply $\delta p_X/2$ in the comoving gauge, and the density perturbation in 
the comoving gauge is given directly by the Poisson equation. Thus we read off the rest frame sound speed of the three-form:
\be \label{cs}
c_S^2 = \frac{\ddot{H}}{\dot{H}}\frac{X}{\dot{X}} -1 = \frac{V_{,XX}X}{V_{,X}} \,,
\ee
where we have used the background relations in the three-form dominated universe. 
For a power law potential $V(X)= X^n$, it results that the speed of sound is a constant given by $c_S^2 = n-1$.
The expression (\ref{cs}) can also be found by noting that in the
rest frame $\alpha=0$ the expressions (\ref{dens}),(\ref{pres}) assume the form 
\bea
\delta \rho_X|_{\alpha=0} & = &  \frac{1}{12X}V_{,X}\delta A^2, \\
\delta p_X|_{\alpha=0} & = &  \frac{1}{12}V_{,XX}\delta A^2.
\eea
Thus the relation of the dynamical sound speed, $c_S^2 = \delta p_X/\delta \rho_X$, to the derivatives of the potential is general and not restricted to three-form dominated background.  

To analyze the behavior of the sound speed in more detail we have considered specific forms of the potential. 

\subsubsection{Power-law potentials, $ V = X^n$}

If the potential is $V = X^n$, the sound speed squared is a constant $c_S^2 = n-1$. 
The quadratic potential resembles a canonical scalar field as its speed
of sound is always the speed of light. With higher order self-interaction terms the fluctuations in the three-form field propagate faster than light, which 
might be seen as a problem. Less controversial is the fact that potentials with $n<1$ are unstable and thus are not viable models of inflation.    
  
\subsubsection{Exponential potentials, $V = \exp\left(-\beta X^2\right)$}

Since the potential $V = \exp\left(-\beta X^2\right)$ can be approximated with a quadratic correction to a constant term, the behavior of the sound speed
is similar to the power-law case,
$c_S^2 = 1-2\beta X^2$.
For negative $\beta$, superluminal propagation is a possibility. For positive $\beta$, however, the value of $X$ is constrained to be $X^2 < 1/2\beta$ to avoid an imaginary speed of sound and consequently  an unstable scenario.

\subsubsection{Ginzburg-Landau potentials, $V=(X^2-C^2)^2$}

Now the expression for the sound speed reads
\be \label{csound}
c_S^2 = 1 + \frac{2X^2}{X^2-C^2}.
\ee
In this case we find that the speed of sound is positive provided that either $X^2 > C^2$ or $X^2 < C^2/3$. Recall from the Section \ref{phase} that in the case $C<\sqrt{2/3}$, the minimum of the potential at $X=\pm C$ is the late time attractor and that at this point the field changes its 
nature from phantom to non-phantom, or vice versa. At the level of the background kinematics, this is perfectly legitimate and can be easily verified by 
explicit solutions. However, scrutiny of the perturbation dynamics reveals that at the phantom divide the sound speed squared diverges and jumps from 
negative to positive infinity or the other way around. Clearly the phantom divide crossing is inhibited in reality for this particular potential.  

The form of the potential is such that it suggest that the origin corresponds to an unstable cosmological constant. This motivates us to consider a case where the 
three-form drives an inflationary period at the local maximum of the potential at $X \approx 0$  which can be seen as a natural initial value. As the field eventually drops to the true minimum, several possibilities may take place. 
If $C > \sqrt{2/3}$ then $X$ is constrained to be $|X| < \sqrt{2/3}$ hence for $c_S^2 > 0$ we must require $C > \sqrt{2}$, and the field will asymptotically approach 
$X = \pm \sqrt{2/3}$. 
In the $C<\sqrt{2/3}$ case, reaching the minimum will lead to instabilities which might be beneficial by generating efficient reheating. 
This possibility might be worth being analyzed in the future.

%%%%%%%%%%%%%%%%%%%%%%%%%%%%%%%%%%%%%%%%%%%%%%%%%%%%%%%%%%%%%

\subsection{Scalar and tensor power spectra from inflation}

As we have seen, it turns out we can describe the scalar fluctuations of the field with only one degree of freedom by exploiting the constraints of the system, 
in particular Eq.~(\ref{const2}). This is due to the symmetries of the FLRW metric. One cannot see directly from the action that one of the four 
degrees of freedom present can be eliminated, as one should see when it is a general property of the theory.
In Bianchi backgrounds, we have many examples of form solutions that are constrained. 
Of course, cosmological perturbations in principle allow all the degrees of freedom present to propagate, but it happens now that the kinetic term has the gauge symmetry which reduces the number of physical degrees of freedom in the absence of the potential.
Even when the potential is turned on, the symmetry is partially efficient. This is because the potential depends only on $A^2$ and the 
spatial components of $A$ are forced to vanish in the FLRW background so their fluctuations $\alpha$ cannot contribute at the linear order to the quadratic
invariant $A^2$ (see Eq.(\ref{square})). Therefore we believe that the absence of the scalar second mode is due to linearization about the isotropic and homogeneous 
solution. Higher order perturbations would thus be interesting to consider, but their study is beyond the present scope. 

To consider quantum fluctuations during inflation, we must find the canonical variable to describe the degree of freedom we have. 
It is conventional to refer to the curvature perturbation $\zeta$ given by
\be \label{zeta}
\zeta = - H \frac{\dot{\phi} + H \phi}{\dot{H}} + \phi \,,
\ee
that evolves as 
\be \label{zetadot}
\dot{\zeta} = -\frac{H}{\dot{H}}c^2_S\frac{1}{a^2} \, \nabla^2\phi,
\ee
as can be verified using (\ref{phievol}). Thus the curvature perturbation is conserved at large scales. By comparing with well-known cases in the 
literature \cite{Mukhanov:1990me}, we can deduce that the canonical degree of freedom is now related to the curvature perturbation as
\be
v \equiv z\zeta, \hspace{1cm} z = \frac{-a\sqrt{-2\dot{H}}}{\kappa c_S H} \,.
\ee
At this point it is convenient to switch to conformal time, and in the remainder of this section a prime will denote derivative wrt to 
conformal time $\tau = \int a \, dt$. It is straightforward to show that the canonical variable now obeys the equation of motion
\be
\label{veq}
v'' - \left( c_S^2\, \nabla^2 + \frac{z''}{z} \right) v =0 \,.
\ee
The action for this variable $v$ could be computed by expanding the action to second order, but this is not necessary.
In the present case the equation of motion fixes the action, though only up to a constant. However, we know the normalization from analogy to 
previous literature \cite{Mukhanov:1990me}. 
Thus we may write
\be
\delta_2 S = \frac{1}{2}\int \left(v'^2-c_S^2\gamma^{ij}v_{,i}v_{,j} + \frac{z''}{z}v^2 \right)\sqrt{\gamma} \, d^4x, 
\ee
where $\gamma_{ij}$ is the metric of spatial sections, which we here assume to be flat for simplicity. 

We proceed quantizing $v$ by promoting the perturbation to an operator and expanding in plane waves, 
\be
\hat{v}(\tau,k) = \int \frac{d^3k}{(2\pi)^{3/2}}\left(\omega_k(\tau) \hat{a}_{\rm k} 
e^{i {\rm k} \cdot {\rm x}}+ \omega_k^*(\tau) 
\hat{a}_{\rm k}^\dagger e^{-i {\rm k} \cdot {\rm x}}\right)  \,, 
\ee
We must now solve the following equation of motion for the wave modes,
\be
\omega_k'' + \left( c_S^2\, k^2 - \frac{z''}{z} \right) \omega_k =0 \,.
\ee
We make the assumptions that the evolution of the universe is power law like with scale factor $a = (-\tau)^p$ with $p = -1/(1-\epsilon)$ which corresponds to 
\be
\epsilon \equiv - \dot{H}/H^2 \,,
\ee
being approximately constant and that
the evolution of the sound speed can be approximated with the power-law form,
\be
\label{sound}
c_S = c_0 (-\tau)^\sigma \,.
\ee
We will see shortly that this is in fact the case for the three-form during slow-roll (super)inflation of this work.
In the case when we have $z''/z \propto \tau^{-2}$,  
the general solution for $\omega_k$ can be written as a sum of the Hankel functions,
and the appropriately normalized solution with positive frequency in the asymptotic past is
\be
\omega_k(\tau) = \frac{1}{2} \sqrt{\frac{\pi}{1+\sigma}} \sqrt{-\tau} H_\nu^{(1)}(x) \,, 
\ee
where $H_\nu^{(1)}$ is the Hankel function of the first kind of order $\nu$, 
with 
\begin{eqnarray}
\label{nu}
\nu &\equiv& \frac{\sqrt{1-4(-z''/z)\tau^2}}{2(1+\sigma)} \,, \\
\label{defx}
x &\equiv& \frac{c_0 k}{1+\sigma} (-\tau)^{1+\sigma} \,.
\end{eqnarray}
Therefore the solution depend on the parameter $\sigma$, and presumably, trough $z''/z$, on $\epsilon$ and other slow roll parameters.

In the long wavelength limit, $x \ll 1$, the behavior of the Hankel function is $H^{(1)}_\nu(x) \rightarrow (i/\pi)\Gamma(\nu)(2/x)^{\nu}$. From this this asymptotic behavior we can calculate the power spectrum of the curvature perturbation on the large scales
\begin{eqnarray}
\mathcal{P}_\zeta & \equiv & \frac{k^3}{2\pi^2} |\zeta_k|^2 \equiv A_\zeta^2 \, \left(\frac{k}{aH}\right)^{n_S-1} \\
&=& \left(\frac{k}{aH}\right)^{3-2\nu} \left(\frac{c_S}{1+\sigma}\right)^{1-2\nu} \frac{c_S \, \kappa^2}{32 \pi^2} \times \nonumber \\
&~& 2^{2\nu-1}\, (1-\epsilon)^{2\nu-1}\, \frac{\Gamma(\nu)^2}{\Gamma(3/2)^2} \, \frac{H^2}{|\epsilon|} \,.
\end{eqnarray}
We can then read the scalar spectral index which is
\be \label{ns}
n_S-1  =  3-2\nu \,.
%\approx -2\epsilon 3 \,.
\ee

Because the three-form does not introduce tensor sources, and the evolution of gravity waves is given by the usual equation (\ref{grav}), we obtain in the 
standard way the spectrum of tensorial perturbations for power law evolution of the universe,
\begin{eqnarray}
\mathcal{P}_T &\equiv& A_T^2 \, \left(\frac{k}{aH}\right)^{n_T} \nonumber \\
&=& \left(\frac{k}{aH}\right)^{3-2\mu} \frac{\kappa^2}{2 \pi^2} \times \nonumber \\
&~& 2^{2\mu-1}\, (1-\epsilon)^{2\mu-1}\, \frac{\Gamma(\mu)^2}{\Gamma(3/2)^2} \, H^2 \,,
\end{eqnarray}
where 
\be
\mu = \frac{1}{2} \sqrt{1-4(-a''/a)\tau^2} \,.
\ee
The tensor spectral index is 
\be
\label{nt}
n_T = 3-2\mu \,.
\ee
We will now compute these quantities in terms of the slow-roll parameter which can be calculated for a given scalar potential.

\subsubsection{Slow-roll (super)inflation}
In order to proceed with the computation of $\nu$ in Eq.~(\ref{nu}) we note that,
\be
z^2 = \frac{2}{\kappa^2}\, \frac{a^2 \epsilon}{c_S^2} \,,
\ee
where the slow-roll parameter $\epsilon$ in our system is written as
\be
\label{epsdef}
\epsilon \equiv -\frac{\dot{H}}{H^2} = \frac{3}{2} \, 
\frac{V_{,X}}{V} \, X \, \left( 1- \frac{\kappa^2}{6}(X_{,N}+3X)^2 \right) \,.
\ee
In what follows we mean inflation when $\epsilon$ is positive and super-inflation when it takes negatives values. 
In order to obtain inflation we do not need slow-roll in the sense of negligible velocity of $X$ (in fact a flat potential leads to de Sitter inflation even if the field is moving, see Eq.~(\ref{eos})). Here, however, 
we are now interested in investigating precisely the slow-roll case.
Then the velocity of $X$ can be neglected, and thus we find that $\epsilon$ can be well approximated by
\be
\label{epsapprox}
\epsilon \approx \frac{3}{2} \, 
\frac{V_{,X}}{V} \, X \, \left( 1- \frac{3}{2} (\kappa X)^2 \right) \,,
\ee
which allows us to immediately determine whether the universe is inflating for a given choice of the scalar potential at a given value of $X$. Using the original equation of motion for $X$ Eq.~(\ref{eomX}) and neglecting the $\ddot{X}$ contribution, it can be found that
\be
\label{dx1}
\kappa X_{,N} = -\frac{V_{,X}}{V} \left(1- \frac{3}{2} (\kappa X)^2 \right)^2 \,.
\ee
This can also be written, using (\ref{epsapprox}) in the useful form 
\be
\label{dx2}
\frac{V_{,X}}{V} X_{,N} = - \frac{4}{9} \frac{\epsilon^2}{(\kappa X)^2}  \,.
\ee
Differentiating $\epsilon$ in Eq.~(\ref{epsdef}) with respect to conformal time and using the equation of motion for $X$ it is obtained
\be
\label{deps}
\epsilon' = 2 \epsilon^2 \eta \, a H \,,
\ee
where we have used Eq.~(\ref{dx2}) and defined
\be
\eta \equiv 1- \frac{2}{9} \frac{\epsilon}{(\kappa X)^2} \left(\frac{V_{,XX} V}{V_{,X}^2} + \frac{V}{V_{,X} X}\right) \,.
\ee
We therefore see that the second term in $\eta$ is suppressed by $\epsilon$ and typically $\eta$ is of order unity. However, Eq.~(\ref{deps}) tells us that $d\ln 
\epsilon/dN = 2 \epsilon \eta$ and suggests that $\epsilon$ can be considered constant when small or equivalently, that the evolution of the universe is power 
law like with scale factor $a = (-\tau)^p$ and $p = -1/(1-\epsilon)$.

Similarly, by differentiating the speed of sound, it is obtained that
\be
c_S' = \epsilon^2 \Theta \, c_S \, aH \,,
\ee
where we have defined 
\be
\Theta \equiv \frac{2}{9} \frac{1}{(\kappa X)^2} \, \frac{V_{,XX}V}{V_{,X}^2} \left(1- \frac{V_{,XXX} V_{,X}}{V_{,XX}^2} - \frac{V_{,X}}{V_{,XX} X}\right) \,.
\ee
For a power law potential we find that $\Theta$ vanishes and therefore $c_S$ is a constant in agreement with what we found earlier. Therefore, when $\epsilon$ is small, both $\epsilon$ and $\Theta$ are nearly constant which enables us to indeed write (\ref{sound}) with 
\be
\sigma = \frac{\epsilon^2 \, \Theta}{1-\epsilon} \,.
\ee
Now Eq.~(\ref{ns}) gives that 
\be
n_S-1 \approx -2\epsilon \left(1+\frac{2}{3} \eta\right) \,.
\ee
We immediately see that near and approaching the fixed point B where $\epsilon$ is negative, the three-form predicts a slightly blue spectral index for curvature perturbations which is disfavored by current observations. This is the case illustrated in Fig.~\ref{x2}. When $\epsilon$ is positive, however, a red tilted spectrum is obtained. Such an example is shown in Fig.~\ref{x1}.

From Eq.~(\ref{nt}) the tensor spectral index is equivalent to standard scalar field inflation
\be
n_T = -2 \epsilon \,.
\ee
Like the curvature spectral index, this is predicted to be slightly blue when the evolution is near and approaching the fixed point B, and red tilted otherwise.
The tensor to scalar ratio is, however, modified. For small $\epsilon$ we have 
\be
r \equiv \frac{A^2_T}{A^2_\zeta} = 16 \, c_S \, |\epsilon| \,. 
\ee
Thus it is in principle possible to distinguish the three-form inflation from scalar field already from the spectra of linear perturbations.

\subsection{Matter perturbations}

The impact of the presence of three-from to the evolution of matter inhomogeneities is considered here. The observationally relevant case is  
dust-like matter (cold dark matter and baryons) in the late-time universe, where anisotropic stresses can be neglected. The conservation equations for matter
then are, in the Fourier space,
\bea
\dot{\delta}_m & = & -\frac{\theta_m}{a}+3\dot{\phi} \,, \\
\frac{1}{a}\dot{\theta}_m & = & -\frac{1}{a}H\theta_m + \frac{k^2}{a^2}\phi \,. \label{mvelo}
\eea 
Combining these gives
\be \label{evol}
\ddot{\delta}_m + 2H\dot{\delta}_m + \frac{k^2}{a^2}\phi = 3\ddot{\phi} + 6H\dot{\phi} \,.
\ee
The effect of the three-form comes thus through the background expansion ($H$ and $a$) and through the coupling of the matter to the 
gravitational potentials. We then specialize to the subhorizon scales. Then the RHS of the previous equation can be neglected. 
We need yet to evaluate the gradient of the gravitational potential. 
At this small-scale limit the perturbed energy constraint is, using (\ref{e00}) and (\ref{dens}),
\be \label{se00}
\frac{k^2}{a^2}\phi = -\frac{1}{2}\rho_m\delta_m + V_{,X}\left(\frac{\dot{K}}{X}\alpha + \alpha_0 + 3X\phi\right) \,. 
\ee
By using the momentum constraint (\ref{e0i}) with (\ref{velo}) and (\ref{mvelo}), we get
\be
V_{,X}\alpha = 2(\dot{\phi}+H\phi) + \frac{1}{2}\rho_m(\dot{\delta}_m - 3\ddot{\phi}) \,. 
\ee
This shows that the contribution from $\alpha$ to the gravitational potential in (\ref{se00}) is suppressed by the $a^2/k^2$, which allows us to 
neglect it at the small scale limit, at least excluding the special case of diverging $\dot{K}/X$. Assuming $(k/a)^2 \gg V_{,X}/X$, we can furthermore 
show that the equations of motion (\ref{a1}) and (\ref{a0}) imply, instead of (\ref{const2}), the algebraic relation
\be
V_{,XX} \alpha_0 = -(3V_{,XX}X+V_{,X})\phi
\ee
between the scalar perturbation and the gravitational potential. Using this in (\ref{se00}) and plugging back to (\ref{evol}), we get
\be \label{evol2}
\ddot{\delta}_m + 2H\dot{\delta}_m = 4\pi G(k,t) \rho_m\delta_m,
\ee
where the scale and time dependent effective gravitational constant is given by
\be \label{newt}
8\pi G(k,t) =  \kappa^2 +\frac{1}{2}\frac{V_{,X}X a^2}{c_S^2 k^2}.
\ee
It is to be expected that the apparent modification is proportional to the slope of the potential, since in this case this slope vanishes
the field reduces to a smooth cosmological constant.  
We also immediately note that if the sound speed squared is large, the small scale limit corresponds to the presence of a smooth component.  
Possible new effects (other than due to modified background expansion) require sound speed less than unity, and the relevant range of scales is the 
intermediate regime between the cosmological horizon and the sound horizon:
\be
H^2 \ll \frac{k^2}{a^2} \ll \frac{H^2}{c_S^2}. 
\ee
The possible scale-dependent signature extends also to larger scales, but from superhorizon scales one typically expects possible observable effects 
only to the largest cosmic variance limited CMB multipoles. For small sound speeds the modifications can extend to nonlinear scales, where we cannot 
trust the expression (\ref{newt}) anymore. In the regime we can trust it, it is in principle possible to constrain some classes of three-form cosmologies 
by using the probes of large scale structure, weak lensing, the integrated Sachs-Wolfe effect and their correlations. 

\section{Formalities}
\label{Formalities}

In the following two subsections we will consider some formal aspects of our simple theory. The reader interested solely in cosmological phenomenology may proceed to the conclusions. 
In particular, we will scrutize the field content of the action (\ref{action}) in terms of a so called St\"uckelberg trick \cite{Ruegg:2003ps} and by performing mappings and duality 
transformations to other theories. One of the aims is to proof 
explicitly the claim made in the introduction that the models can be seen as a novel description of scalar field models in some specific situations but however 
for most cases it is not (even formally) equivalent to a scalar theory. This point, as well as some clarifications of our 
terminology will be useful to make here,
since many different conventions (and some confusions) exist in the literature. 
%The reader interested more in the cosmology of the model or already 
%familiar with the relations between antisymmetric tensor models of different rank is invited to proceed to the next section.

\subsection{Gauge invariance and stability}

The antisymmetrized gradient of the three-form gauge potential term in Eq.~(\ref{action}) is gauge-invariant under transformation 
%$A_{\alpha\beta\gamma} \rightarrow A_{\alpha\beta\gamma} + \nabla_{[\alpha}\Delta_{\beta\gamma]}$. 
$A \rightarrow A + [\nabla \Delta]$, where $\Delta$ is an arbitrary two-form. However, the potential in the action
(\ref{action}) breaks this symmetry. This can be seen to result in extra degrees of freedom in the model, analogous 
to the appearance of longitudinal polarization of the massive photon in the Proca theory. To make this explicit,
one may introduce a St\"uckelberg form $\Sigma$ in such a way that for a redefined $A = \tilde{A} + 4[\nabla \Sigma]$, 
the Lagrangian 
\be
\mathcal{L} = -\frac{1}{48}F^2(\tilde{A}) - V\left((\tilde{A}+F(\Sigma))^2\right)
\ee
is then manifestly invariant under the gauge transformations
\be
\tilde{A} \rightarrow \tilde{A} + [\nabla \Delta], \hspace{1cm} \Sigma \rightarrow \Sigma - \Delta/4.
\ee
In the $\Sigma=0$ gauge we recover our original Lagrangian. If there is a gauge where the gradient of $\Delta$ is orthogonal to $\tilde{A}$, and 
if we assume we can expand around a background solution given by constant $A$, we can write
\be \label{stuck}
\mathcal{L} = \mathcal{L} - V'(\tilde{A}^2) F^2(\Sigma),
\ee
making transparent the appearance of the extra two-form degree of freedom. This also seems to imply that the extra degree of freedom becomes a ghost when 
$V'(A^2)<0$. We remind that this is also exactly the condition for the three-form to violate
the null energy condition in FLRW background, i.e. to become phantom-like with equation of state less than $-1$. However, investigating the stability of the
field in more detail by considering inhomogeneous and anisotropic fluctuations of the field and taking into account their backreaction due to coupling to the 
metric, we find that the conditions for the stability of the canonical degrees of freedom can be more subtle than the naive implication of (\ref{stuck}).   
In fact, the behavior of the field depends on the second derivative of the potential since the propagation speed of the physical fluctuations turned out 
to be given by (equivalently to the expression (\ref{cs}) in terms of the comoving field)
\be   
c_S^2 = 1+ 2 \frac{V''(A^2)A^2}{V'(A^2)} \,.
\ee
The field, even if phantom, can be stable at least classically. However, divergences tend to occur at the ''phantom divide'' when $V'(A^2)$ crosses zero.
Similar phenomena, linking classical singularities and quantum no-ghost conditions have been observed in other models 
\cite{Koivisto:2006ai,Koivisto:2008xfa,Jimenez:2009ai}. The conditions for the possibility 
of a viable phantom crossing is now the following: twice differentiable $V(x)$ exists for positive $x$ in such a way that $V'(x)$ changes sign at $x=x_0$, 
and $x_0(\ln(V'(x_0)))'$ is finite.  
 
This can easily change if nonminimal couplings are introduced, but in the present paper we consider only minimal couplings.    

%%%%%%%%%%%%%%%%%%%%%%%%%%%%%%%%%%%%%%%%%%%%%%%%%%%%%%%%%%%%%%%%%%
\subsection{Duality and equivalence of theories}
\label{duality}

\begin{figure}
\includegraphics[width=8.5cm]{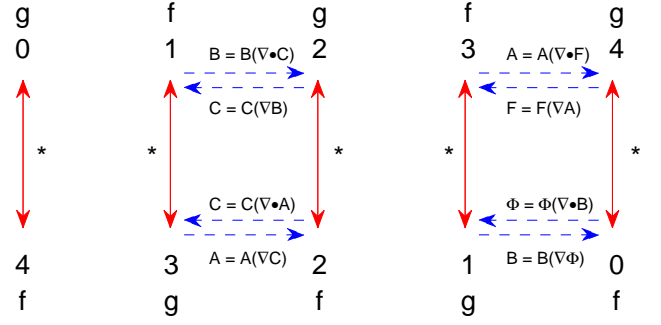}
\caption{\label{dual} A summary of the chains of equivalences between forms. In the case of canonic form, $f$ refers
to a model with the Faraday kinetic term with an action of the form (\ref{faraday}) and $x$ to a model with the dual (gauge-fixing) kinetic term with
an action of the form (\ref{gaugefixing}). The Hodge 
duality operation is vertical movement denoted with a star, and a horizontal movement is a change variables for which schematic formulas are given. The first group 
in the figure, consisting of the form-form and dual, is trivial. 
The second group, which depicts the chain (\ref{line1}), would correspond to a starting point of having the gauge-fixing kinetic term for the three-form.
The third group depicts the chain (\ref{line2}), corresponding to the case whose generalization we consider in this paper. 
} 
\end{figure}

In this subsection we will discuss some equivalences between n-form models. First we introduce a parent Lagrangian (\ref{parent}) 
\be \label{parent}
\mathcal{L}_p = \frac{1}{48}F^2 - \frac{1}{6}A \, \nabla\cdot F - V(A^2) \,,
\ee
which can be rewritten in terms of a Faraday form
\be
\label{faraday}
\mathcal{L}_f = f\left( F^2(x) \right) - V(x^2) \,,
\ee
where $f$ and $V$ are arbitrary functions and $x$ is a n-form, which
describes our starting point 
(\ref{action}) for $x = A$. We will also show that also a four-form can emerge from it, i.e., we can rewrite the parent Lagrangian in terms of gauge-fixing terms only
\be
\label{gaugefixing}
\mathcal{L}_g = g\left( (\nabla \cdot x)^2 \right) - U(x^2) \,,
\ee
where $g$ and $U$ are functions. 
Next we show how the Hodge dual of the parent Lagrangian (\ref{lag}) can give rise to a vector or a scalar field description. Finally, the chain of equivalences is summarized in the diagram (\ref{dual}). Dualities in the case of nonminimal gravity couplings have been discussed in \cite{Germani:2009iq}.

Starting with the parent Lagrangian (\ref{parent}), where 
$F$ is an independent four-form and solving its equation of motion, one gets $F=-4 [\nabla A]$. By plugging this back into (\ref{parent}) 
one obtains the original action (\ref{action}). 

One may also integrate out the three-form and obtain a theory for the four-form $F$ as follows.
Varying with respect to $A$ gives us the equation of motion
\be \label{4form1}
-\frac{1}{6}\nabla\cdot F + 2V'(A^2)A = 0 \,,
\ee
implying
\be \label{4form2}
Y \equiv (\nabla\cdot F)^2 = 144 \, \left(V'(A^2)\right)^2 \, A^2 \,.
\ee
If we now plug $A$ from (\ref{4form1}) and the solution $A^2(Y)$ (\ref{4form2}) into (\ref{parent}), we get a dynamical four-form theory:
\be \label{4form3}
\mathcal{L} = \frac{1}{48}F^2 - \frac{1}{72}\, \frac{Y}{V'\left(A^2(Y)\right)}-V\left(A^2(Y)\right).
\ee
The Lagrangian is now written in a gauge-fixing form (\ref{gaugefixing}) with $x = F$.
The Faraday form constructed from a four-form is of course trivial, $F(F)=0$, and static four-forms contribute only a constant. 
As mentioned in the introduction, this has been employed in attempts to solve the cosmological constant problem 
\cite{Hawking:1984hk,Turok:1998he}.
Four-form formulation of $f(R)$ gravity \cite{Koivisto:2009sd} and
dark energy from promoting the Levi-Civita symbol into dynamical form have been considered recently \cite{Gupta:2009jy}.

The Hodge dual of the three-form is a vector $(\ast A)$. Writing the parent Lagrangian (\ref{parent}) in terms of the dual forms
\be \label{duals}
F = \epsilon (\ast F) \equiv  \epsilon \Phi \,, \hspace{1cm}
A = \epsilon (\ast A) \equiv  \epsilon B,
\ee
where $\Phi$ is a scalar field and $B$ is a vector,
we obtain
\be \label{lag}
\mathcal{L} = -\frac{1}{2}\Phi^2-B \, \nabla\Phi - V\left(-B^2/6\right) \,.
\ee
Now the equation of motion for $\Phi$ is simply that $\Phi=\nabla\cdot B$, and replacing this back gives us the self-coupled vector 
theory 
\be \label{1form}
\mathcal{L} = \frac{1}{2}\left(\nabla\cdot B\right)^2 - V\left(-B^2/6\right) \,,
\ee
in a gauge-fixing form as in (\ref{gaugefixing}) with $x = B$.
Recently the cosmological significance of the Maxwell theory supplemented with the gauge-fixing term like in (\ref{1form}) has been 
considered, and very interestingly it has been found that the gauge-fixing term results in an effective cosmological 
constant in the curved background (or almost constant, since it fluctuates) while the theory in the Minkowski limit reduces to 
standard electromagnetism \cite{Jimenez:2008nm,Jimenez:2009sv}. Such a Maxwell theory is dual to three-form having both the 
$F^2(A)$ and $(\nabla\cdot A)^2$, but presently we confine to include only the canonical term yielding only the gauge-fixing 
term for the corresponding vector.  

Finally, we may integrate out the vector from (\ref{lag}) to obtain a scalar field theory. The Euler-Lagrange equation $B$ is
\be
\nabla\Phi = \frac{1}{3} \, V'\left(-B^2/6\right)B,
\ee
implying
\be \label{0eq}
9 (\nabla\Phi)^2 = \left[V'\left(-B^2/6\right)\right]^2B^2.
\ee
Similarly as with the four-form, we assume these equations are invertible, and then write the Lagrangian (\ref{lag}) in terms of them as
\be \label{0form}
\mathcal{L} = -\frac{1}{2}\Phi^2 - \frac{Y_S}{3 V'\left(-B^2(Y_S)/6\right)}-V\left(-B^2(Y_S)/6\right),
\ee
where now $Y_S \equiv (\nabla\Phi)^2$. The dual Lagrangian is now written in a Faraday type (\ref{faraday}) with $x = \Phi$. 
This completes our task of deriving the equivalent reformulations we mentioned in the introduction. 

It is useful to note that only two possibilities of canonical forms exist in four dimensions. Any such form is either a vector or a dual vector. 
The former is dual to a three-form, and can be rewritten as two-form which is self-dual. The latter can be seen as scalar field, thus dual to a four-form and 
consequently rewritable as a three-form. Schematically, the chains of equivalences can be written as 
\bea
f1 & \leftrightarrow &  g3  =  f2 \leftrightarrow g2 = f1, \label{line1} \\
g1 & = & f0 \leftrightarrow g4  =  f3 \leftrightarrow g1 \,. \label{line2}
\eea
Here $f{\rm n}$ denotes an n-form described by a Faraday type Lagrangian (\ref{faraday}), and 
$g{\rm n}$ an n-form described by gauge-fixing terms type Lagrangian (\ref{gaugefixing}). Duality in the Hodge sense is indicated with $\leftrightarrow$, and $=$ means equivalence between Lagrangians by change of variables. This is also shown in diagram \ref{dual}.

The case we consider in the present paper belongs to the chain (\ref{line2}) or the third group in diagram \ref{dual}. Thus it seems we 
are, at the same time, discussing four types of theories: 
\begin{itemize}
\item a model of vector with self-interactions and a gauge-fixing type term, Eq.(\ref{1form}),
\item a K-essence type scalar field model Eq.(\ref{0form}), 
\item a four-form with dynamics due to a nonstandard kinetic term, Eq.(\ref{4form3}),
\item a canonical three-form, Eq.(\ref{action}).
\end{itemize}
These equivalences are valid due to the possibility of rewriting the four degrees of freedom in the three-form as a vector. Furthermore, if the vector is exact, i.e. expressible as
a gradient of a scalar, the model can be reduced to a scalar field, equivalently a four-form.

However, we should immediately mention that these equivalences break down in many cases. 
In the above derivations, this breakdown occurs when there are no real solutions to equations (\ref{4form2}) or (\ref{0eq}).
The example potential of most interest in the present study of the action (\ref{action}) is of the form of a displaced power-law 
$V(A^2) = (A^2/6-C^2)^2$, since that 
is the case that includes most of the features of the possible dynamics. In particular, having a non-monotonic first derivative allows the field to dynamically change its nature. 
A scalar field formulation, however, seems not to be available. Another simple class of potentials we considered is the exponential potential. As shown in
appendix \ref{ap_s}, this class does not seem to admit a scalar field formulation either. Even in some cases where the equivalence formally holds, the formulation in the three-form language is much more transparent.

\subsection{Generalizations}

One might consider extensions of the simple and minimal action (\ref{action}). Since, due to the self-interactions of the gauge potential $A$, there is no 
gauge invariance, the gauge invariance cannot be used as a criterion for the action. However, gauge-invariant nonlinear terms can be considered and 
motivated by quantum corrections such as in nonlinear electrodynamics \cite{Novello:1979ik,Novello:2006ng}. Here a phenomenological motivation for such terms was 
briefly mentioned at the end of 
section \ref{Applications}. When going to curved spacetime, it is known that terms involving contractions of $F$ and 
the Riemann and Ricci tensors arise as leading order quantum corrections \cite{Drummond:1979pp}. Models obtained by the inclusion of the 
gauge-invariance preserving forms like $f(R,G)F^2$, where $f(R,G)$ is some function of the curvature scalar or Gauss-Bonnet term, have been applied in 
cosmology in cases of Maxwell or Yang-Mills fields \cite{Bamba:2008ja,Sadeghi:2009pu,Bamba:2008be}. 

As mentioned, the potentials break gauge-invariance and one can allow more gauge-invariance breaking terms. These could still be quadratic. In 
particular, a general quadratic second order Lagrangian for a three-form would read,
\bea 
\label{g_action}
-\mathcal{L} & = & \alpha_1 F^2(A) + \alpha_2 (\nabla\cdot A)^2 + \left(\alpha_3 R + \frac{1}{2}m^2 \right) A^2 
\nonumber \\ & + &  \alpha_4 A\cdot Ric \cdot A + 
\alpha_5A\cdot\cdot (\mathcal{R}\cdot\cdot A),
\eea
where $\mathcal{R}$ is the Riemann and $Ric$ the Ricci tensor. There are thus six coefficients to specify such a theory. If the principle of minimal coupling 
is kept as a guide as in the present study, only three are left. A particular combination of the coefficients, keeping only $\alpha_2=0$ but fixed 
values for the rest $\alpha_i$ results in a 
scalar-field like equation of motion for the comoving field in a FLRW background. For this reason such a model has been considered as generalization of the 
vector inflation scheme \cite{Germani:2009iq,Kobayashi:2009hj}. In future work, it would be interesting to study the consistency constrains of 
nonzero $\alpha_i$, in particular the implications of the gauge-fixing term given by a nonzero $\alpha_2$.
Many of the parameter possible combinations in the general action (\ref{g_action}) are probably excluded due to appearance of ghosts and instabilities.
In addition, Solar system experiments can be used to constrain the nonminimal couplings to gravity, $\alpha_3$, $\alpha_4$ and $\alpha_5$. To our knowledge these 
issues have not been addressed in the case of three-form, whereas vector models
have been extensively studied (see Ref.~\cite{Will:1972zz} and references mentioned in the introduction section \ref{introduction}).
Finally, since the Maxwell field allows a dual description as a three-form, one may contemplate whether it is possible that nonminimal couplings can lead to variations of fundamental parameters such as the fine structure constant.  

\section{Conclusions}
\label{conclusions}

We considered the evolution of the universe in the presence of three-forms. We assumed a canonical and minimally coupled action taking into account the possibility of self-interactions of the form field.
Then a form with three differing spatial indices is compatible with an isotropic and homogeneous cosmological background. It turns out that such form, despite its
canonical form, quite generically violates the usual energy conditions. The strong energy condition is violated when 
\be \label{cond1}
V_{,X}X < \frac{4}{3}\left(\frac{1}{2}(\dot{X}+ 3HX)^2 + V\right ) \,\, 
\Rightarrow \,\, w_X < -\frac{1}{3}.
\ee
This happens quite easily. Slow roll is not required for accelerating behavior, only that $V_{,X}X$ is not large compared to the energy density. The null energy 
condition is broken and the field behaves as phantom when
\be \label{cond2}
V_{,X}X<0 \,\,\Rightarrow \,\, w_X < -1.
\ee
Generally stability problems appear in crossing the phantom divide, but they might be overcome by nonminimal couplings.
In the absence of a potential, the effect of the field reduces simply to generating a cosmological constant.
Thus the three-form seems a very suitable culprit for the accelerating expansion which is believed to take place both at an early stage and at a late 
stage of the history of the universe. One notes that phantom divide crossing is possible for simple forms of the potential, such as the Landau-Ginzburg form that we 
took as our main example.

We performed a phase space analysis of the model and found three distinct fixed points: matter domination, a potential extremum and a peculiar fixed point 
corresponding
to a kinetic domination of the three-form (and a potential domination of the dual scalar field Eq.~(\ref{0form}) in the special cases that such a scalar field 
exists). The latter 
two fixed points describe de Sitter spaces, and their nature and stability depends on the potential. The mathematical properties of the system proved nontrivial, 
and forced us to go up to third order in the perturbations for some forms of the potential. This tempts one to investigate the model in the bifurcation theory framework. On the physical side,
one of the de Sitter fixed points is always an attractor. Thus the fixed points are interesting for inflationary and dark energy applications. For the latter, 
one wishes to find initial conditions independent dynamics before the acceleration in order to reduce the fine-tuning. 
Previously it was shown that scaling and tracking solutions exits and the requirements for 
these to last for a large number of e-folds can be quantified \cite{Koivisto:2009ew}. It is then clear that one may realize a rich variety of background dynamics 
using a single three-form as the energy source.

The three-form fluctuations were then investigated. They were parameterized in terms of two scalar and two vector modes (whose form 
was motivated by the dual vector).  
The latter are (as usual) phenomenologically less interesting. However, it is worthwhile to note that in general these degrees of freedom also exist. 
It turned out that one may describe the scalar perturbations with one effective degree of freedom, since the constraints allowed to eliminate 
one of the two fields parameterizing the scalar fluctuations.   
The presence of the three-form can modify the evolution of scalar fluctuations in matter distribution. If the three-form 
sound speed is sufficiently less than unity, one expects possibly observable effects for the large scale structure, weak lensing, the integrated 
Sachs-Wolfe effect 
and their correlations. The impact of three-form fluctuations can be quantified by introducing an effective gravitational constant defined by 
\be \label{cond3}
8\pi G(k,t) = \kappa^2 +\frac{1}{2}\frac{V^2_{,X}a^2}{V_{,XX}k^2} \,.
\ee
During a three-form driven inflation the nontrivial dynamics can also lead to sound speed dependent possibly detectable signatures. We identified the 
canonical degrees of freedom and quantized them. A slow-roll parameterization was reconsidered for this new case and a very convenient way of studying the quantum 
generation of 
perturbations near a de Sitter fixed point was developed. The spectral indices of scalar and tensor perturbations have a easily computable dependence on the form of
the potential through the sound speed,
\be \label{cond4}
c_S^2 = \frac{V_{,XX}X}{V_{,X}} \,.
\ee
The tensor to scalar ratio is modified directly by this quantity. 

The four formulas (\ref{cond1})--(\ref{cond4}) summarize how the shape of the potential determines the nature of the field and thus the background dynamics and 
properties of fluctuations for a given model. To conclude, we have shown that these objects, also present in string theory, can give rise to viable cosmological 
scenarios with potentially observable signatures distinct from standard single scalar field models.

\appendix

\section{The scalar field formulation}
\label{ap_s}

In this Appendix we derive explicitly the scalar field formulation corresponding to a simple exponential potential and a displaced power-law potential.
The main conclusion to be drawn is that the scalar field formulation of 
these models is typically non-trivial. 
The mapping of these models 
is not a bijection, but the defining equations have multiple solutions and branches. Furthermore, some of the solutions become complex, so mapping 
becomes ill-defined or ceases to exist. 

Consider the potential $V(A^2) = V_0 (6A^2-C^2)^2$, where we have introduced the irrelevant rescaling by the factor of $6$ just simplify some formulas. 
Then Eq.(\ref{0eq}) reads
\be
\frac{1}{4}(\nabla \Phi)^2 = 4V_0^2(6A^2-C^2)^2 A^2 \,.
\ee
This a third order equation for the square of the form: in general three solutions exist. For the sake of explicitness, we pick
up the solution which seems to lead to the simplest form of the Lagrangian. This Lagrangian is then
\bea
\mathcal{L} & = &  -  \frac{1}{144} V_0 \left(\frac{ 2^{2/3}8 V_0^2 C^4}{f((\nabla \Phi)^2)} + 20C^2 + \frac{\sqrt[3]{2}f((\nabla 
\Phi)^2)}{V_0^2}\right)^2  \nonumber \\ 
   & - & \frac{3 V_0 (\nabla \Phi)^2}{\frac{2^{2/3}8 C^4 V_0^4}{f((\nabla \Phi)^2)}+20 C^2 V_0^2 + f((\nabla \Phi)^2)} + \frac{1}{2}m^2\Phi^2,
\eea
where
\bea
f((\nabla \Phi)^2) & = & \Big[-32 C^6 V^6 + 27V_0^4(\nabla \Phi)^2 + \\ \nonumber 
& + & 3 \sqrt{3} \sqrt{V_0^8 (\nabla \Phi)^2 \left(27 (\nabla \Phi)^2-64 C^6 V_0^2\right)}\Big]^\frac{1}{3}. 
\eea
This is a noncanonical field indeed.

Consider the exponential potential $V(A^2) = V_0e^{-\lambda 6A^2}$. Then Eq.(\ref{0eq}) reads
\be
\frac{1}{4}(\nabla \Phi)^2 = \lambda^2V_0^2e^{-2\lambda A^2}A^2.
\ee
The solution can be written in terms of the Lambert's $W$-function as
\be
-A^2 = \frac{1}{2\lambda}W(x),
\ee
where we defined
\be
x= \frac{-(\nabla \Phi)^2}{2\lambda V_0^2}.
\ee
Lambert's $W$-function has multiple branches. It is real $W(x) \in \mathbb{R}$, when the argument $x \ge 1/e$. The scalar field Lagrangian would then read
\be
\mathcal{L} = -V_0\left(x e^{\frac{1}{2\lambda}W(x)} - e^{-\frac{1}{2\lambda}W(x)}\right) - \frac{1}{2}\Phi^2.
\ee
It seems that in general such a model is not well defined. In any case it is clear that the formulation as a canonic form with an exponential potential is
considerably more tractable than this formulation. 

\section{About the dual vector field}
\label{ap_v}

Here we state some facts concerning the dual vector field $\ast A$ appearing in (\ref{duals}). We obtain it by the change of variables, and it can be written 
explicitly using the indices as
\be \label{vector}
\ast A_\mu = \frac{1}{6}\epsilon_{\alpha\beta\gamma\mu}A^{\alpha\beta\gamma} \,.  
\ee
Since the equations come out in equivalent form regardless whether we begin with $A$ or with $\ast A$, we find no benefit in switching between alternative 
descriptions. However, one exception appears in our parameterization of the three-form perturbation in Eqs.~(\ref{0ij}) and (\ref{ijk}), 
\bea
A_{0ij} & = & a^2(t)\epsilon_{ijk}(\alpha_{,k} + \alpha_k) \label{0ij2}, \\
A_{ijk} & = & a^3(t)\epsilon_{ijk}(X(t)+\alpha_0). \label{ijk2} 
\eea
To justify calling the components of scalar perturbations as if they were components of a vector, one may employ the gauge transformations 
Eqs.~(\ref{g1})--(\ref{g3}) to see that they indeed transform as if they form a vector. A more direct way to see this is to note that the dual vector is just
\bea
\ast A_0 &=& \frac{1}{a^3}\left(X + \alpha_0 + 6X\phi\right),  \\
\ast A_i &=& \frac{1}{a^3}\left(\alpha_{,i} + \alpha_i\right).
\eea
One could have absorbed $6X\phi$ into the definition of $\alpha_0$.

\begin{acknowledgments}

We acknowledge n-formal discussions with Jos\'e Beltr\'an Jim\'enez, David Mota, David Mulryne, Sami Nurmi and Cyril Pitrou.  
The authors are supported by Deutsche Forschungsgemeinschaft, project TRR33.
This work was initiated at the workshop ''New horizons for modern cosmology'' at the Galileo Galilei Institute in Florence; we would like to thank the
institute for hospitality.
 
\end{acknowledgments}

\bibliography{refs}

\begin{thebibliography}{94}
\expandafter\ifx\csname natexlab\endcsname\relax\def\natexlab#1{#1}\fi
\expandafter\ifx\csname bibnamefont\endcsname\relax
  \def\bibnamefont#1{#1}\fi
\expandafter\ifx\csname bibfnamefont\endcsname\relax
  \def\bibfnamefont#1{#1}\fi
\expandafter\ifx\csname citenamefont\endcsname\relax
  \def\citenamefont#1{#1}\fi
\expandafter\ifx\csname url\endcsname\relax
  \def\url#1{\texttt{#1}}\fi
\expandafter\ifx\csname urlprefix\endcsname\relax\def\urlprefix{URL }\fi
\providecommand{\bibinfo}[2]{#2}
\providecommand{\eprint}[2][]{\url{#2}}

\bibitem[{\citenamefont{Nordstrom}(1914)}]{Nordstrom:1988fi}
\bibinfo{author}{\bibfnamefont{G.}~\bibnamefont{Nordstrom}},
  \bibinfo{journal}{Phys. Z.} \textbf{\bibinfo{volume}{15}},
  \bibinfo{pages}{504} (\bibinfo{year}{1914}), \eprint{physics/0702221}.

\bibitem[{\citenamefont{Brans and Dicke}(1961)}]{Brans:1961sx}
\bibinfo{author}{\bibfnamefont{C.}~\bibnamefont{Brans}} \bibnamefont{and}
  \bibinfo{author}{\bibfnamefont{R.~H.} \bibnamefont{Dicke}},
  \bibinfo{journal}{Phys. Rev.} \textbf{\bibinfo{volume}{124}},
  \bibinfo{pages}{925} (\bibinfo{year}{1961}).

\bibitem[{\citenamefont{Starobinsky}(1980)}]{Starobinsky:1980te}
\bibinfo{author}{\bibfnamefont{A.~A.} \bibnamefont{Starobinsky}},
  \bibinfo{journal}{Phys. Lett.} \textbf{\bibinfo{volume}{B91}},
  \bibinfo{pages}{99} (\bibinfo{year}{1980}).

\bibitem[{\citenamefont{Wetterich}(1988)}]{Wetterich:1987fm}
\bibinfo{author}{\bibfnamefont{C.}~\bibnamefont{Wetterich}},
  \bibinfo{journal}{Nucl. Phys.} \textbf{\bibinfo{volume}{B302}},
  \bibinfo{pages}{668} (\bibinfo{year}{1988}).

\bibitem[{\citenamefont{Peebles and Ratra}(1988)}]{Peebles:1987ek}
\bibinfo{author}{\bibfnamefont{P.~J.~E.} \bibnamefont{Peebles}}
  \bibnamefont{and} \bibinfo{author}{\bibfnamefont{B.}~\bibnamefont{Ratra}},
  \bibinfo{journal}{Astrophys. J.} \textbf{\bibinfo{volume}{325}},
  \bibinfo{pages}{L17} (\bibinfo{year}{1988}).

\bibitem[{\citenamefont{Amendola}(2000)}]{Amendola:1999er}
\bibinfo{author}{\bibfnamefont{L.}~\bibnamefont{Amendola}},
  \bibinfo{journal}{Phys. Rev.} \textbf{\bibinfo{volume}{D62}},
  \bibinfo{pages}{043511} (\bibinfo{year}{2000}), \eprint{astro-ph/9908023}.

\bibitem[{\citenamefont{Koivisto and
  Mota}(2007{\natexlab{a}})}]{Koivisto:2006xf}
\bibinfo{author}{\bibfnamefont{T.}~\bibnamefont{Koivisto}} \bibnamefont{and}
  \bibinfo{author}{\bibfnamefont{D.~F.} \bibnamefont{Mota}},
  \bibinfo{journal}{Phys. Lett.} \textbf{\bibinfo{volume}{B644}},
  \bibinfo{pages}{104} (\bibinfo{year}{2007}{\natexlab{a}}),
  \eprint{astro-ph/0606078}.

\bibitem[{\citenamefont{Bassett et~al.}(2006)\citenamefont{Bassett, Tsujikawa,
  and Wands}}]{Bassett:2005xm}
\bibinfo{author}{\bibfnamefont{B.~A.} \bibnamefont{Bassett}},
  \bibinfo{author}{\bibfnamefont{S.}~\bibnamefont{Tsujikawa}},
  \bibnamefont{and} \bibinfo{author}{\bibfnamefont{D.}~\bibnamefont{Wands}},
  \bibinfo{journal}{Rev. Mod. Phys.} \textbf{\bibinfo{volume}{78}},
  \bibinfo{pages}{537} (\bibinfo{year}{2006}), \eprint{astro-ph/0507632}.

\bibitem[{\citenamefont{Copeland et~al.}(2006)\citenamefont{Copeland, Sami, and
  Tsujikawa}}]{Copeland:2006wr}
\bibinfo{author}{\bibfnamefont{E.~J.} \bibnamefont{Copeland}},
  \bibinfo{author}{\bibfnamefont{M.}~\bibnamefont{Sami}}, \bibnamefont{and}
  \bibinfo{author}{\bibfnamefont{S.}~\bibnamefont{Tsujikawa}},
  \bibinfo{journal}{Int. J. Mod. Phys.} \textbf{\bibinfo{volume}{D15}},
  \bibinfo{pages}{1753} (\bibinfo{year}{2006}), \eprint{hep-th/0603057}.

\bibitem[{\citenamefont{Ford}(1989)}]{Ford:1989me}
\bibinfo{author}{\bibfnamefont{L.~H.} \bibnamefont{Ford}},
  \bibinfo{journal}{Phys. Rev.} \textbf{\bibinfo{volume}{D40}},
  \bibinfo{pages}{967} (\bibinfo{year}{1989}).

\bibitem[{\citenamefont{Koivisto and
  Mota}(2008{\natexlab{a}})}]{Koivisto:2008xf}
\bibinfo{author}{\bibfnamefont{T.~S.} \bibnamefont{Koivisto}} \bibnamefont{and}
  \bibinfo{author}{\bibfnamefont{D.~F.} \bibnamefont{Mota}},
  \bibinfo{journal}{JCAP} \textbf{\bibinfo{volume}{0808}}, \bibinfo{pages}{021}
  (\bibinfo{year}{2008}{\natexlab{a}}), \eprint{0805.4229}.

\bibitem[{\citenamefont{Golovnev et~al.}(2008)\citenamefont{Golovnev, Mukhanov,
  and Vanchurin}}]{Golovnev:2008cf}
\bibinfo{author}{\bibfnamefont{A.}~\bibnamefont{Golovnev}},
  \bibinfo{author}{\bibfnamefont{V.}~\bibnamefont{Mukhanov}}, \bibnamefont{and}
  \bibinfo{author}{\bibfnamefont{V.}~\bibnamefont{Vanchurin}},
  \bibinfo{journal}{JCAP} \textbf{\bibinfo{volume}{0806}}, \bibinfo{pages}{009}
  (\bibinfo{year}{2008}), \eprint{0802.2068}.

\bibitem[{\citenamefont{Himmetoglu et~al.}(2009)\citenamefont{Himmetoglu,
  Contaldi, and Peloso}}]{Himmetoglu:2008zp}
\bibinfo{author}{\bibfnamefont{B.}~\bibnamefont{Himmetoglu}},
  \bibinfo{author}{\bibfnamefont{C.~R.} \bibnamefont{Contaldi}},
  \bibnamefont{and} \bibinfo{author}{\bibfnamefont{M.}~\bibnamefont{Peloso}},
  \bibinfo{journal}{Phys. Rev. Lett.} \textbf{\bibinfo{volume}{102}},
  \bibinfo{pages}{111301} (\bibinfo{year}{2009}), \eprint{0809.2779}.

\bibitem[{\citenamefont{Karciauskas et~al.}(2008)\citenamefont{Karciauskas,
  Dimopoulos, and Lyth}}]{Karciauskas:2008bc}
\bibinfo{author}{\bibfnamefont{M.}~\bibnamefont{Karciauskas}},
  \bibinfo{author}{\bibfnamefont{K.}~\bibnamefont{Dimopoulos}},
  \bibnamefont{and} \bibinfo{author}{\bibfnamefont{D.~H.} \bibnamefont{Lyth}}
  (\bibinfo{year}{2008}), \eprint{0812.0264}.

\bibitem[{\citenamefont{Golovnev and Vanchurin}(2009)}]{Golovnev:2009ks}
\bibinfo{author}{\bibfnamefont{A.}~\bibnamefont{Golovnev}} \bibnamefont{and}
  \bibinfo{author}{\bibfnamefont{V.}~\bibnamefont{Vanchurin}},
  \bibinfo{journal}{Phys. Rev.} \textbf{\bibinfo{volume}{D79}},
  \bibinfo{pages}{103524} (\bibinfo{year}{2009}), \eprint{0903.2977}.

\bibitem[{\citenamefont{Watanabe et~al.}(2009)\citenamefont{Watanabe, Kanno,
  and Soda}}]{Watanabe:2009ct}
\bibinfo{author}{\bibfnamefont{M.-a.} \bibnamefont{Watanabe}},
  \bibinfo{author}{\bibfnamefont{S.}~\bibnamefont{Kanno}}, \bibnamefont{and}
  \bibinfo{author}{\bibfnamefont{J.}~\bibnamefont{Soda}}
  (\bibinfo{year}{2009}), \eprint{0902.2833}.

\bibitem[{\citenamefont{Yokoyama and Soda}(2008)}]{Yokoyama:2008xw}
\bibinfo{author}{\bibfnamefont{S.}~\bibnamefont{Yokoyama}} \bibnamefont{and}
  \bibinfo{author}{\bibfnamefont{J.}~\bibnamefont{Soda}},
  \bibinfo{journal}{JCAP} \textbf{\bibinfo{volume}{0808}}, \bibinfo{pages}{005}
  (\bibinfo{year}{2008}), \eprint{0805.4265}.

\bibitem[{\citenamefont{Dimopoulos
  et~al.}(2009{\natexlab{a}})\citenamefont{Dimopoulos, Karciauskas, Lyth, and
  Rodriguez}}]{Dimopoulos:2008yv}
\bibinfo{author}{\bibfnamefont{K.}~\bibnamefont{Dimopoulos}},
  \bibinfo{author}{\bibfnamefont{M.}~\bibnamefont{Karciauskas}},
  \bibinfo{author}{\bibfnamefont{D.~H.} \bibnamefont{Lyth}}, \bibnamefont{and}
  \bibinfo{author}{\bibfnamefont{Y.}~\bibnamefont{Rodriguez}},
  \bibinfo{journal}{JCAP} \textbf{\bibinfo{volume}{0905}}, \bibinfo{pages}{013}
  (\bibinfo{year}{2009}{\natexlab{a}}), \eprint{0809.1055}.

\bibitem[{\citenamefont{Dimopoulos
  et~al.}(2009{\natexlab{b}})\citenamefont{Dimopoulos, Karciauskas, and
  Wagstaff}}]{Dimopoulos:2009am}
\bibinfo{author}{\bibfnamefont{K.}~\bibnamefont{Dimopoulos}},
  \bibinfo{author}{\bibfnamefont{M.}~\bibnamefont{Karciauskas}},
  \bibnamefont{and} \bibinfo{author}{\bibfnamefont{J.~M.}
  \bibnamefont{Wagstaff}} (\bibinfo{year}{2009}{\natexlab{b}}),
  \eprint{0907.1838}.

\bibitem[{\citenamefont{Kiselev}(2004)}]{Kiselev:2004py}
\bibinfo{author}{\bibfnamefont{V.~V.} \bibnamefont{Kiselev}},
  \bibinfo{journal}{Class. Quant. Grav.} \textbf{\bibinfo{volume}{21}},
  \bibinfo{pages}{3323} (\bibinfo{year}{2004}), \eprint{gr-qc/0402095}.

\bibitem[{\citenamefont{Armendariz-Picon}(2004)}]{ArmendarizPicon:2004pm}
\bibinfo{author}{\bibfnamefont{C.}~\bibnamefont{Armendariz-Picon}},
  \bibinfo{journal}{JCAP} \textbf{\bibinfo{volume}{0407}}, \bibinfo{pages}{007}
  (\bibinfo{year}{2004}), \eprint{astro-ph/0405267}.

\bibitem[{\citenamefont{Koivisto and Mota}(2006)}]{Koivisto:2005mm}
\bibinfo{author}{\bibfnamefont{T.}~\bibnamefont{Koivisto}} \bibnamefont{and}
  \bibinfo{author}{\bibfnamefont{D.~F.} \bibnamefont{Mota}},
  \bibinfo{journal}{Phys. Rev.} \textbf{\bibinfo{volume}{D73}},
  \bibinfo{pages}{083502} (\bibinfo{year}{2006}), \eprint{astro-ph/0512135}.

\bibitem[{\citenamefont{Boehmer and Harko}(2007)}]{Boehmer:2007qa}
\bibinfo{author}{\bibfnamefont{C.~G.} \bibnamefont{Boehmer}} \bibnamefont{and}
  \bibinfo{author}{\bibfnamefont{T.}~\bibnamefont{Harko}},
  \bibinfo{journal}{Eur. Phys. J.} \textbf{\bibinfo{volume}{C50}},
  \bibinfo{pages}{423} (\bibinfo{year}{2007}), \eprint{gr-qc/0701029}.

\bibitem[{\citenamefont{Koivisto and
  Mota}(2008{\natexlab{b}})}]{Koivisto:2007bp}
\bibinfo{author}{\bibfnamefont{T.}~\bibnamefont{Koivisto}} \bibnamefont{and}
  \bibinfo{author}{\bibfnamefont{D.~F.} \bibnamefont{Mota}},
  \bibinfo{journal}{Astrophys. J.} \textbf{\bibinfo{volume}{679}},
  \bibinfo{pages}{1} (\bibinfo{year}{2008}{\natexlab{b}}), \eprint{0707.0279}.

\bibitem[{\citenamefont{Wei and Cai}(2006)}]{Wei:2006tn}
\bibinfo{author}{\bibfnamefont{H.}~\bibnamefont{Wei}} \bibnamefont{and}
  \bibinfo{author}{\bibfnamefont{R.-G.} \bibnamefont{Cai}},
  \bibinfo{journal}{Phys. Rev.} \textbf{\bibinfo{volume}{D73}},
  \bibinfo{pages}{083002} (\bibinfo{year}{2006}), \eprint{astro-ph/0603052}.

\bibitem[{\citenamefont{Koivisto and
  Mota}(2008{\natexlab{c}})}]{Koivisto:2008ig}
\bibinfo{author}{\bibfnamefont{T.}~\bibnamefont{Koivisto}} \bibnamefont{and}
  \bibinfo{author}{\bibfnamefont{D.~F.} \bibnamefont{Mota}},
  \bibinfo{journal}{JCAP} \textbf{\bibinfo{volume}{0806}}, \bibinfo{pages}{018}
  (\bibinfo{year}{2008}{\natexlab{c}}), \eprint{0801.3676}.

\bibitem[{\citenamefont{Jimenez and
  Maroto}(2009{\natexlab{a}})}]{Jimenez:2009ai}
\bibinfo{author}{\bibfnamefont{J.~B.} \bibnamefont{Jimenez}} \bibnamefont{and}
  \bibinfo{author}{\bibfnamefont{A.~L.} \bibnamefont{Maroto}}
  (\bibinfo{year}{2009}{\natexlab{a}}), \eprint{0905.1245}.

\bibitem[{\citenamefont{Mota et~al.}(2007)\citenamefont{Mota, Kristiansen,
  Koivisto, and Groeneboom}}]{Mota:2007sz}
\bibinfo{author}{\bibfnamefont{D.~F.} \bibnamefont{Mota}},
  \bibinfo{author}{\bibfnamefont{J.~R.} \bibnamefont{Kristiansen}},
  \bibinfo{author}{\bibfnamefont{T.}~\bibnamefont{Koivisto}}, \bibnamefont{and}
  \bibinfo{author}{\bibfnamefont{N.~E.} \bibnamefont{Groeneboom}},
  \bibinfo{journal}{Mon. Not. Roy. Astron. Soc.}
  \textbf{\bibinfo{volume}{382}}, \bibinfo{pages}{793} (\bibinfo{year}{2007}),
  \eprint{0708.0830}.

\bibitem[{\citenamefont{Dulaney et~al.}(2008)\citenamefont{Dulaney, Gresham,
  and Wise}}]{Dulaney:2008ph}
\bibinfo{author}{\bibfnamefont{T.~R.} \bibnamefont{Dulaney}},
  \bibinfo{author}{\bibfnamefont{M.~I.} \bibnamefont{Gresham}},
  \bibnamefont{and} \bibinfo{author}{\bibfnamefont{M.~B.} \bibnamefont{Wise}},
  \bibinfo{journal}{Phys. Rev.} \textbf{\bibinfo{volume}{D77}},
  \bibinfo{pages}{083510} (\bibinfo{year}{2008}), \eprint{0801.2950}.

\bibitem[{\citenamefont{Germani and
  Kehagias}(2009{\natexlab{a}})}]{Germani:2009iq}
\bibinfo{author}{\bibfnamefont{C.}~\bibnamefont{Germani}} \bibnamefont{and}
  \bibinfo{author}{\bibfnamefont{A.}~\bibnamefont{Kehagias}},
  \bibinfo{journal}{JCAP} \textbf{\bibinfo{volume}{0903}}, \bibinfo{pages}{028}
  (\bibinfo{year}{2009}{\natexlab{a}}), \eprint{0902.3667}.

\bibitem[{\citenamefont{Kobayashi and Yokoyama}(2009)}]{Kobayashi:2009hj}
\bibinfo{author}{\bibfnamefont{T.}~\bibnamefont{Kobayashi}} \bibnamefont{and}
  \bibinfo{author}{\bibfnamefont{S.}~\bibnamefont{Yokoyama}},
  \bibinfo{journal}{JCAP} \textbf{\bibinfo{volume}{0905}}, \bibinfo{pages}{004}
  (\bibinfo{year}{2009}), \eprint{0903.2769}.

\bibitem[{\citenamefont{Koivisto et~al.}(2009)\citenamefont{Koivisto, Mota, and
  Pitrou}}]{Koivisto:2009sd}
\bibinfo{author}{\bibfnamefont{T.~S.} \bibnamefont{Koivisto}},
  \bibinfo{author}{\bibfnamefont{D.~F.} \bibnamefont{Mota}}, \bibnamefont{and}
  \bibinfo{author}{\bibfnamefont{C.}~\bibnamefont{Pitrou}}
  (\bibinfo{year}{2009}), \eprint{0903.4158}.

\bibitem[{\citenamefont{Saha}(2006)}]{Saha:2006iu}
\bibinfo{author}{\bibfnamefont{B.}~\bibnamefont{Saha}}, \bibinfo{journal}{Phys.
  Rev.} \textbf{\bibinfo{volume}{D74}}, \bibinfo{pages}{124030}
  (\bibinfo{year}{2006}).

\bibitem[{\citenamefont{Boehmer and Mota}(2008)}]{Boehmer:2007ut}
\bibinfo{author}{\bibfnamefont{C.~G.} \bibnamefont{Boehmer}} \bibnamefont{and}
  \bibinfo{author}{\bibfnamefont{D.~F.} \bibnamefont{Mota}},
  \bibinfo{journal}{Phys. Lett.} \textbf{\bibinfo{volume}{B663}},
  \bibinfo{pages}{168} (\bibinfo{year}{2008}), \eprint{0710.2003}.

\bibitem[{\citenamefont{Zhao and Zhang}(2006)}]{Zhao:2005bu}
\bibinfo{author}{\bibfnamefont{W.}~\bibnamefont{Zhao}} \bibnamefont{and}
  \bibinfo{author}{\bibfnamefont{Y.}~\bibnamefont{Zhang}},
  \bibinfo{journal}{Class. Quant. Grav.} \textbf{\bibinfo{volume}{23}},
  \bibinfo{pages}{3405} (\bibinfo{year}{2006}), \eprint{astro-ph/0510356}.

\bibitem[{\citenamefont{Bamba et~al.}(2008)\citenamefont{Bamba, Nojiri, and
  Odintsov}}]{Bamba:2008xa}
\bibinfo{author}{\bibfnamefont{K.}~\bibnamefont{Bamba}},
  \bibinfo{author}{\bibfnamefont{S.}~\bibnamefont{Nojiri}}, \bibnamefont{and}
  \bibinfo{author}{\bibfnamefont{S.~D.} \bibnamefont{Odintsov}},
  \bibinfo{journal}{Phys. Rev.} \textbf{\bibinfo{volume}{D77}},
  \bibinfo{pages}{123532} (\bibinfo{year}{2008}), \eprint{0803.3384}.

\bibitem[{\citenamefont{Copeland et~al.}(1995)\citenamefont{Copeland, Lahiri,
  and Wands}}]{Copeland:1994km}
\bibinfo{author}{\bibfnamefont{E.~J.} \bibnamefont{Copeland}},
  \bibinfo{author}{\bibfnamefont{A.}~\bibnamefont{Lahiri}}, \bibnamefont{and}
  \bibinfo{author}{\bibfnamefont{D.}~\bibnamefont{Wands}},
  \bibinfo{journal}{Phys. Rev.} \textbf{\bibinfo{volume}{D51}},
  \bibinfo{pages}{1569} (\bibinfo{year}{1995}), \eprint{hep-th/9410136}.

\bibitem[{\citenamefont{Lukas et~al.}(1997)\citenamefont{Lukas, Ovrut, and
  Waldram}}]{Lukas:1996iq}
\bibinfo{author}{\bibfnamefont{A.}~\bibnamefont{Lukas}},
  \bibinfo{author}{\bibfnamefont{B.~A.} \bibnamefont{Ovrut}}, \bibnamefont{and}
  \bibinfo{author}{\bibfnamefont{D.}~\bibnamefont{Waldram}},
  \bibinfo{journal}{Nucl. Phys.} \textbf{\bibinfo{volume}{B495}},
  \bibinfo{pages}{365} (\bibinfo{year}{1997}), \eprint{hep-th/9610238}.

\bibitem[{\citenamefont{Gasperini and Veneziano}(1999)}]{Gasperini:1998bm}
\bibinfo{author}{\bibfnamefont{M.}~\bibnamefont{Gasperini}} \bibnamefont{and}
  \bibinfo{author}{\bibfnamefont{G.}~\bibnamefont{Veneziano}},
  \bibinfo{journal}{Phys. Rev.} \textbf{\bibinfo{volume}{D59}},
  \bibinfo{pages}{043503} (\bibinfo{year}{1999}), \eprint{hep-ph/9806327}.

\bibitem[{\citenamefont{Bilic et~al.}(2005)\citenamefont{Bilic, Tupper, and
  Viollier}}]{Bilic:2005zk}
\bibinfo{author}{\bibfnamefont{N.}~\bibnamefont{Bilic}},
  \bibinfo{author}{\bibfnamefont{G.~B.} \bibnamefont{Tupper}},
  \bibnamefont{and} \bibinfo{author}{\bibfnamefont{R.~D.}
  \bibnamefont{Viollier}} (\bibinfo{year}{2005}), \eprint{hep-th/0504082}.

\bibitem[{\citenamefont{Bilic et~al.}(2007)\citenamefont{Bilic, Tupper, and
  Viollier}}]{Bilic:2006cp}
\bibinfo{author}{\bibfnamefont{N.}~\bibnamefont{Bilic}},
  \bibinfo{author}{\bibfnamefont{G.~B.} \bibnamefont{Tupper}},
  \bibnamefont{and} \bibinfo{author}{\bibfnamefont{R.~D.}
  \bibnamefont{Viollier}}, \bibinfo{journal}{J. Phys.}
  \textbf{\bibinfo{volume}{A40}}, \bibinfo{pages}{6877} (\bibinfo{year}{2007}),
  \eprint{gr-qc/0610104}.

\bibitem[{\citenamefont{De~Risi}(2008)}]{DeRisi:2007dn}
\bibinfo{author}{\bibfnamefont{G.}~\bibnamefont{De~Risi}},
  \bibinfo{journal}{Phys. Rev.} \textbf{\bibinfo{volume}{D77}},
  \bibinfo{pages}{044030} (\bibinfo{year}{2008}), \eprint{0711.3781}.

\bibitem[{\citenamefont{Moffat}(1995)}]{Moffat:1994hv}
\bibinfo{author}{\bibfnamefont{J.~W.} \bibnamefont{Moffat}},
  \bibinfo{journal}{Phys. Lett.} \textbf{\bibinfo{volume}{B355}},
  \bibinfo{pages}{447} (\bibinfo{year}{1995}), \eprint{gr-qc/9411006}.

\bibitem[{\citenamefont{Damour et~al.}(1993)\citenamefont{Damour, Deser, and
  McCarthy}}]{Damour:1992bt}
\bibinfo{author}{\bibfnamefont{T.}~\bibnamefont{Damour}},
  \bibinfo{author}{\bibfnamefont{S.}~\bibnamefont{Deser}}, \bibnamefont{and}
  \bibinfo{author}{\bibfnamefont{J.~G.} \bibnamefont{McCarthy}},
  \bibinfo{journal}{Phys. Rev.} \textbf{\bibinfo{volume}{D47}},
  \bibinfo{pages}{1541} (\bibinfo{year}{1993}), \eprint{gr-qc/9207003}.

\bibitem[{\citenamefont{Prokopec and Valkenburg}(2006)}]{Prokopec:2005fb}
\bibinfo{author}{\bibfnamefont{T.}~\bibnamefont{Prokopec}} \bibnamefont{and}
  \bibinfo{author}{\bibfnamefont{W.}~\bibnamefont{Valkenburg}},
  \bibinfo{journal}{Phys. Lett.} \textbf{\bibinfo{volume}{B636}},
  \bibinfo{pages}{1} (\bibinfo{year}{2006}), \eprint{astro-ph/0503289}.

\bibitem[{\citenamefont{Janssen and Prokopec}(2006)}]{Janssen:2006nn}
\bibinfo{author}{\bibfnamefont{T.}~\bibnamefont{Janssen}} \bibnamefont{and}
  \bibinfo{author}{\bibfnamefont{T.}~\bibnamefont{Prokopec}},
  \bibinfo{journal}{Class. Quant. Grav.} \textbf{\bibinfo{volume}{23}},
  \bibinfo{pages}{4967} (\bibinfo{year}{2006}), \eprint{gr-qc/0604094}.

\bibitem[{\citenamefont{Gupta}(2009)}]{Gupta:2009jy}
\bibinfo{author}{\bibfnamefont{P.~D.} \bibnamefont{Gupta}}
  (\bibinfo{year}{2009}), \eprint{0905.1621}.

\bibitem[{\citenamefont{Guendelman and Kaganovich}(2008)}]{Guendelman:2008ms}
\bibinfo{author}{\bibfnamefont{E.~I.} \bibnamefont{Guendelman}}
  \bibnamefont{and} \bibinfo{author}{\bibfnamefont{A.~B.}
  \bibnamefont{Kaganovich}}, \bibinfo{journal}{Class. Quant. Grav.}
  \textbf{\bibinfo{volume}{25}}, \bibinfo{pages}{235015}
  (\bibinfo{year}{2008}), \eprint{0804.1278}.

\bibitem[{\citenamefont{Guendelman and Kaganovich}(2007)}]{Guendelman:2006af}
\bibinfo{author}{\bibfnamefont{E.~I.} \bibnamefont{Guendelman}}
  \bibnamefont{and} \bibinfo{author}{\bibfnamefont{A.~B.}
  \bibnamefont{Kaganovich}}, \bibinfo{journal}{Phys. Rev.}
  \textbf{\bibinfo{volume}{D75}}, \bibinfo{pages}{083505}
  (\bibinfo{year}{2007}), \eprint{gr-qc/0607111}.

\bibitem[{\citenamefont{Guendelman}(1999)}]{Guendelman:1999qt}
\bibinfo{author}{\bibfnamefont{E.~I.} \bibnamefont{Guendelman}},
  \bibinfo{journal}{Mod. Phys. Lett.} \textbf{\bibinfo{volume}{A14}},
  \bibinfo{pages}{1043} (\bibinfo{year}{1999}), \eprint{gr-qc/9901017}.

\bibitem[{\citenamefont{Guendelman and Katz}(2003)}]{Guendelman:2002js}
\bibinfo{author}{\bibfnamefont{E.~I.} \bibnamefont{Guendelman}}
  \bibnamefont{and} \bibinfo{author}{\bibfnamefont{O.}~\bibnamefont{Katz}},
  \bibinfo{journal}{Class. Quant. Grav.} \textbf{\bibinfo{volume}{20}},
  \bibinfo{pages}{1715} (\bibinfo{year}{2003}), \eprint{gr-qc/0211095}.

\bibitem[{\citenamefont{Armendariz-Picon and
  Duvvuri}(2004)}]{ArmendarizPicon:2003qw}
\bibinfo{author}{\bibfnamefont{C.}~\bibnamefont{Armendariz-Picon}}
  \bibnamefont{and} \bibinfo{author}{\bibfnamefont{V.}~\bibnamefont{Duvvuri}},
  \bibinfo{journal}{Class. Quant. Grav.} \textbf{\bibinfo{volume}{21}},
  \bibinfo{pages}{2011} (\bibinfo{year}{2004}), \eprint{hep-th/0305237}.

\bibitem[{\citenamefont{Gunther et~al.}(2003)\citenamefont{Gunther, Moniz, and
  Zhuk}}]{Gunther:2003zn}
\bibinfo{author}{\bibfnamefont{U.}~\bibnamefont{Gunther}},
  \bibinfo{author}{\bibfnamefont{P.}~\bibnamefont{Moniz}}, \bibnamefont{and}
  \bibinfo{author}{\bibfnamefont{A.}~\bibnamefont{Zhuk}},
  \bibinfo{journal}{Phys. Rev.} \textbf{\bibinfo{volume}{D68}},
  \bibinfo{pages}{044010} (\bibinfo{year}{2003}), \eprint{hep-th/0303023}.

\bibitem[{\citenamefont{Saidov and Zhuk}(2007)}]{Saidov:2006xr}
\bibinfo{author}{\bibfnamefont{T.}~\bibnamefont{Saidov}} \bibnamefont{and}
  \bibinfo{author}{\bibfnamefont{A.}~\bibnamefont{Zhuk}},
  \bibinfo{journal}{Phys. Rev.} \textbf{\bibinfo{volume}{D75}},
  \bibinfo{pages}{084037} (\bibinfo{year}{2007}), \eprint{hep-th/0612217}.

\bibitem[{\citenamefont{Jacobson and Mattingly}(2001)}]{Jacobson:2000xp}
\bibinfo{author}{\bibfnamefont{T.}~\bibnamefont{Jacobson}} \bibnamefont{and}
  \bibinfo{author}{\bibfnamefont{D.}~\bibnamefont{Mattingly}},
  \bibinfo{journal}{Phys. Rev.} \textbf{\bibinfo{volume}{D64}},
  \bibinfo{pages}{024028} (\bibinfo{year}{2001}), \eprint{gr-qc/0007031}.

\bibitem[{\citenamefont{Kostelecky and Lehnert}(2001)}]{Kostelecky:2000mm}
\bibinfo{author}{\bibfnamefont{V.~A.} \bibnamefont{Kostelecky}}
  \bibnamefont{and} \bibinfo{author}{\bibfnamefont{R.}~\bibnamefont{Lehnert}},
  \bibinfo{journal}{Phys. Rev.} \textbf{\bibinfo{volume}{D63}},
  \bibinfo{pages}{065008} (\bibinfo{year}{2001}), \eprint{hep-th/0012060}.

\bibitem[{\citenamefont{Li et~al.}(2008)\citenamefont{Li, Fonseca~Mota, and
  Barrow}}]{Li:2007vz}
\bibinfo{author}{\bibfnamefont{B.}~\bibnamefont{Li}},
  \bibinfo{author}{\bibfnamefont{D.}~\bibnamefont{Fonseca~Mota}},
  \bibnamefont{and} \bibinfo{author}{\bibfnamefont{J.~D.}
  \bibnamefont{Barrow}}, \bibinfo{journal}{Phys. Rev.}
  \textbf{\bibinfo{volume}{D77}}, \bibinfo{pages}{024032}
  (\bibinfo{year}{2008}), \eprint{0709.4581}.

\bibitem[{\citenamefont{Zuntz et~al.}(2008)\citenamefont{Zuntz, Ferreira, and
  Zlosnik}}]{Zuntz:2008zz}
\bibinfo{author}{\bibfnamefont{J.~A.} \bibnamefont{Zuntz}},
  \bibinfo{author}{\bibfnamefont{P.~G.} \bibnamefont{Ferreira}},
  \bibnamefont{and} \bibinfo{author}{\bibfnamefont{T.~G.}
  \bibnamefont{Zlosnik}}, \bibinfo{journal}{Phys. Rev. Lett.}
  \textbf{\bibinfo{volume}{101}}, \bibinfo{pages}{261102}
  (\bibinfo{year}{2008}), \eprint{0808.1824}.

\bibitem[{\citenamefont{Carroll et~al.}(2009)\citenamefont{Carroll, Dulaney,
  Gresham, and Tam}}]{Carroll:2009en}
\bibinfo{author}{\bibfnamefont{S.~M.} \bibnamefont{Carroll}},
  \bibinfo{author}{\bibfnamefont{T.~R.} \bibnamefont{Dulaney}},
  \bibinfo{author}{\bibfnamefont{M.~I.} \bibnamefont{Gresham}},
  \bibnamefont{and} \bibinfo{author}{\bibfnamefont{H.}~\bibnamefont{Tam}},
  \bibinfo{journal}{Phys. Rev.} \textbf{\bibinfo{volume}{D79}},
  \bibinfo{pages}{065012} (\bibinfo{year}{2009}), \eprint{0812.1050}.

\bibitem[{\citenamefont{Campanelli}(2009)}]{Campanelli:2009tk}
\bibinfo{author}{\bibfnamefont{L.}~\bibnamefont{Campanelli}}
  (\bibinfo{year}{2009}), \eprint{0907.3703}.

\bibitem[{\citenamefont{de~Oliveira-Costa
  et~al.}(2004)\citenamefont{de~Oliveira-Costa, Tegmark, Zaldarriaga, and
  Hamilton}}]{OliveiraCosta:2003pu}
\bibinfo{author}{\bibfnamefont{A.}~\bibnamefont{de~Oliveira-Costa}},
  \bibinfo{author}{\bibfnamefont{M.}~\bibnamefont{Tegmark}},
  \bibinfo{author}{\bibfnamefont{M.}~\bibnamefont{Zaldarriaga}},
  \bibnamefont{and} \bibinfo{author}{\bibfnamefont{A.}~\bibnamefont{Hamilton}},
  \bibinfo{journal}{Phys. Rev.} \textbf{\bibinfo{volume}{D69}},
  \bibinfo{pages}{063516} (\bibinfo{year}{2004}), \eprint{astro-ph/0307282}.

\bibitem[{\citenamefont{Schwarz et~al.}(2004)\citenamefont{Schwarz, Starkman,
  Huterer, and Copi}}]{Schwarz:2004gk}
\bibinfo{author}{\bibfnamefont{D.~J.} \bibnamefont{Schwarz}},
  \bibinfo{author}{\bibfnamefont{G.~D.} \bibnamefont{Starkman}},
  \bibinfo{author}{\bibfnamefont{D.}~\bibnamefont{Huterer}}, \bibnamefont{and}
  \bibinfo{author}{\bibfnamefont{C.~J.} \bibnamefont{Copi}},
  \bibinfo{journal}{Phys. Rev. Lett.} \textbf{\bibinfo{volume}{93}},
  \bibinfo{pages}{221301} (\bibinfo{year}{2004}), \eprint{astro-ph/0403353}.

\bibitem[{\citenamefont{Eriksen et~al.}(2005)\citenamefont{Eriksen, Banday,
  Gorski, and Lilje}}]{Eriksen:2004iu}
\bibinfo{author}{\bibfnamefont{H.~K.} \bibnamefont{Eriksen}},
  \bibinfo{author}{\bibfnamefont{A.~J.} \bibnamefont{Banday}},
  \bibinfo{author}{\bibfnamefont{K.~M.} \bibnamefont{Gorski}},
  \bibnamefont{and} \bibinfo{author}{\bibfnamefont{P.~B.} \bibnamefont{Lilje}},
  \bibinfo{journal}{Astrophys. J.} \textbf{\bibinfo{volume}{622}},
  \bibinfo{pages}{58} (\bibinfo{year}{2005}), \eprint{astro-ph/0407271}.

\bibitem[{\citenamefont{Samal et~al.}(2008)\citenamefont{Samal, Saha, Jain, and
  Ralston}}]{Samal:2007nw}
\bibinfo{author}{\bibfnamefont{P.~K.} \bibnamefont{Samal}},
  \bibinfo{author}{\bibfnamefont{R.}~\bibnamefont{Saha}},
  \bibinfo{author}{\bibfnamefont{P.}~\bibnamefont{Jain}}, \bibnamefont{and}
  \bibinfo{author}{\bibfnamefont{J.~P.} \bibnamefont{Ralston}},
  \bibinfo{journal}{Mon. Not. Roy. Astron. Soc.}
  \textbf{\bibinfo{volume}{385}}, \bibinfo{pages}{1718} (\bibinfo{year}{2008}),
  \eprint{0708.2816}.

\bibitem[{\citenamefont{Hoftuft et~al.}(2009)}]{Hoftuft:2009rq}
\bibinfo{author}{\bibfnamefont{J.}~\bibnamefont{Hoftuft}} \bibnamefont{et~al.},
  \bibinfo{journal}{Astrophys. J.} \textbf{\bibinfo{volume}{699}},
  \bibinfo{pages}{985} (\bibinfo{year}{2009}), \eprint{0903.1229}.

\bibitem[{\citenamefont{Blomqvist et~al.}(2008)\citenamefont{Blomqvist,
  Mortsell, and Nobili}}]{Blomqvist:2008ud}
\bibinfo{author}{\bibfnamefont{M.}~\bibnamefont{Blomqvist}},
  \bibinfo{author}{\bibfnamefont{E.}~\bibnamefont{Mortsell}}, \bibnamefont{and}
  \bibinfo{author}{\bibfnamefont{S.}~\bibnamefont{Nobili}},
  \bibinfo{journal}{JCAP} \textbf{\bibinfo{volume}{0806}}, \bibinfo{pages}{027}
  (\bibinfo{year}{2008}), \eprint{0806.0496}.

\bibitem[{\citenamefont{Cooray et~al.}(2008)\citenamefont{Cooray, Holz, and
  Caldwell}}]{Cooray:2008qn}
\bibinfo{author}{\bibfnamefont{A.~R.} \bibnamefont{Cooray}},
  \bibinfo{author}{\bibfnamefont{D.~E.} \bibnamefont{Holz}}, \bibnamefont{and}
  \bibinfo{author}{\bibfnamefont{R.}~\bibnamefont{Caldwell}}
  (\bibinfo{year}{2008}), \eprint{0812.0376}.

\bibitem[{\citenamefont{Koivisto}(2007)}]{Koivisto:2007sq}
\bibinfo{author}{\bibfnamefont{T.}~\bibnamefont{Koivisto}},
  \bibinfo{journal}{Phys. Rev.} \textbf{\bibinfo{volume}{D76}},
  \bibinfo{pages}{043527} (\bibinfo{year}{2007}), \eprint{0706.0974}.

\bibitem[{\citenamefont{Quercellini
  et~al.}(2009{\natexlab{a}})\citenamefont{Quercellini, Quartin, and
  Amendola}}]{Quercellini:2008ty}
\bibinfo{author}{\bibfnamefont{C.}~\bibnamefont{Quercellini}},
  \bibinfo{author}{\bibfnamefont{M.}~\bibnamefont{Quartin}}, \bibnamefont{and}
  \bibinfo{author}{\bibfnamefont{L.}~\bibnamefont{Amendola}},
  \bibinfo{journal}{Phys. Rev. Lett.} \textbf{\bibinfo{volume}{102}},
  \bibinfo{pages}{151302} (\bibinfo{year}{2009}{\natexlab{a}}),
  \eprint{0809.3675}.

\bibitem[{\citenamefont{Quercellini
  et~al.}(2009{\natexlab{b}})\citenamefont{Quercellini, Cabella, Amendola,
  Quartin, and Balbi}}]{Quercellini:2009ni}
\bibinfo{author}{\bibfnamefont{C.}~\bibnamefont{Quercellini}},
  \bibinfo{author}{\bibfnamefont{P.}~\bibnamefont{Cabella}},
  \bibinfo{author}{\bibfnamefont{L.}~\bibnamefont{Amendola}},
  \bibinfo{author}{\bibfnamefont{M.}~\bibnamefont{Quartin}}, \bibnamefont{and}
  \bibinfo{author}{\bibfnamefont{A.}~\bibnamefont{Balbi}}
  (\bibinfo{year}{2009}{\natexlab{b}}), \eprint{0905.4853}.

\bibitem[{\citenamefont{Duff and van Nieuwenhuizen}(1980)}]{Duff:1980qv}
\bibinfo{author}{\bibfnamefont{M.~J.} \bibnamefont{Duff}} \bibnamefont{and}
  \bibinfo{author}{\bibfnamefont{P.}~\bibnamefont{van Nieuwenhuizen}},
  \bibinfo{journal}{Phys. Lett.} \textbf{\bibinfo{volume}{B94}},
  \bibinfo{pages}{179} (\bibinfo{year}{1980}).

\bibitem[{\citenamefont{Hawking}(1984)}]{Hawking:1984hk}
\bibinfo{author}{\bibfnamefont{S.~W.} \bibnamefont{Hawking}},
  \bibinfo{journal}{Phys. Lett.} \textbf{\bibinfo{volume}{B134}},
  \bibinfo{pages}{403} (\bibinfo{year}{1984}).

\bibitem[{\citenamefont{Turok and Hawking}(1998)}]{Turok:1998he}
\bibinfo{author}{\bibfnamefont{N.}~\bibnamefont{Turok}} \bibnamefont{and}
  \bibinfo{author}{\bibfnamefont{S.~W.} \bibnamefont{Hawking}},
  \bibinfo{journal}{Phys. Lett.} \textbf{\bibinfo{volume}{B432}},
  \bibinfo{pages}{271} (\bibinfo{year}{1998}), \eprint{hep-th/9803156}.

\bibitem[{\citenamefont{Koivisto and Nunes}(2009)}]{Koivisto:2009ew}
\bibinfo{author}{\bibfnamefont{T.~S.} \bibnamefont{Koivisto}} \bibnamefont{and}
  \bibinfo{author}{\bibfnamefont{N.~J.} \bibnamefont{Nunes}}
  (\bibinfo{year}{2009}), \eprint{0907.3883}.

\bibitem[{\citenamefont{Gubser}(2000)}]{Gubser:2000vg}
\bibinfo{author}{\bibfnamefont{S.~S.} \bibnamefont{Gubser}}
  (\bibinfo{year}{2000}), \eprint{hep-th/0010010}.

\bibitem[{\citenamefont{Frey and Mazumdar}(2003)}]{Frey:2002qc}
\bibinfo{author}{\bibfnamefont{A.~R.} \bibnamefont{Frey}} \bibnamefont{and}
  \bibinfo{author}{\bibfnamefont{A.}~\bibnamefont{Mazumdar}},
  \bibinfo{journal}{Phys. Rev.} \textbf{\bibinfo{volume}{D67}},
  \bibinfo{pages}{046006} (\bibinfo{year}{2003}), \eprint{hep-th/0210254}.

\bibitem[{\citenamefont{Germani and
  Kehagias}(2009{\natexlab{b}})}]{Germani:2009gg}
\bibinfo{author}{\bibfnamefont{C.}~\bibnamefont{Germani}} \bibnamefont{and}
  \bibinfo{author}{\bibfnamefont{A.}~\bibnamefont{Kehagias}}
  (\bibinfo{year}{2009}{\natexlab{b}}), \eprint{0908.0001}.

\bibitem[{\citenamefont{Armendariz-Picon
  et~al.}(1999)\citenamefont{Armendariz-Picon, Damour, and
  Mukhanov}}]{ArmendarizPicon:1999rj}
\bibinfo{author}{\bibfnamefont{C.}~\bibnamefont{Armendariz-Picon}},
  \bibinfo{author}{\bibfnamefont{T.}~\bibnamefont{Damour}}, \bibnamefont{and}
  \bibinfo{author}{\bibfnamefont{V.~F.} \bibnamefont{Mukhanov}},
  \bibinfo{journal}{Phys. Lett.} \textbf{\bibinfo{volume}{B458}},
  \bibinfo{pages}{209} (\bibinfo{year}{1999}), \eprint{hep-th/9904075}.

\bibitem[{\citenamefont{Gruzinov}(2004)}]{Gruzinov:2004rq}
\bibinfo{author}{\bibfnamefont{A.}~\bibnamefont{Gruzinov}}
  (\bibinfo{year}{2004}), \eprint{astro-ph/0401520}.

\bibitem[{\citenamefont{{Turner}}(1983)}]{1983PhRvD..28.1243T}
\bibinfo{author}{\bibfnamefont{M.~S.} \bibnamefont{{Turner}}},
  \bibinfo{journal}{\prd} \textbf{\bibinfo{volume}{28}}, \bibinfo{pages}{1243}
  (\bibinfo{year}{1983}).

\bibitem[{\citenamefont{Mukhanov et~al.}(1992)\citenamefont{Mukhanov, Feldman,
  and Brandenberger}}]{Mukhanov:1990me}
\bibinfo{author}{\bibfnamefont{V.~F.} \bibnamefont{Mukhanov}},
  \bibinfo{author}{\bibfnamefont{H.~A.} \bibnamefont{Feldman}},
  \bibnamefont{and} \bibinfo{author}{\bibfnamefont{R.~H.}
  \bibnamefont{Brandenberger}}, \bibinfo{journal}{Phys. Rept.}
  \textbf{\bibinfo{volume}{215}}, \bibinfo{pages}{203} (\bibinfo{year}{1992}).

\bibitem[{\citenamefont{Ruegg and Ruiz-Altaba}(2004)}]{Ruegg:2003ps}
\bibinfo{author}{\bibfnamefont{H.}~\bibnamefont{Ruegg}} \bibnamefont{and}
  \bibinfo{author}{\bibfnamefont{M.}~\bibnamefont{Ruiz-Altaba}},
  \bibinfo{journal}{Int. J. Mod. Phys.} \textbf{\bibinfo{volume}{A19}},
  \bibinfo{pages}{3265} (\bibinfo{year}{2004}), \eprint{hep-th/0304245}.

\bibitem[{\citenamefont{Koivisto and
  Mota}(2007{\natexlab{b}})}]{Koivisto:2006ai}
\bibinfo{author}{\bibfnamefont{T.}~\bibnamefont{Koivisto}} \bibnamefont{and}
  \bibinfo{author}{\bibfnamefont{D.~F.} \bibnamefont{Mota}},
  \bibinfo{journal}{Phys. Rev.} \textbf{\bibinfo{volume}{D75}},
  \bibinfo{pages}{023518} (\bibinfo{year}{2007}{\natexlab{b}}),
  \eprint{hep-th/0609155}.

\bibitem[{\citenamefont{Koivisto}(2008)}]{Koivisto:2008xfa}
\bibinfo{author}{\bibfnamefont{T.}~\bibnamefont{Koivisto}},
  \bibinfo{journal}{Phys. Rev.} \textbf{\bibinfo{volume}{D77}},
  \bibinfo{pages}{123513} (\bibinfo{year}{2008}), \eprint{0803.3399}.

\bibitem[{\citenamefont{Jimenez and
  Maroto}(2009{\natexlab{b}})}]{Jimenez:2008nm}
\bibinfo{author}{\bibfnamefont{J.~B.} \bibnamefont{Jimenez}} \bibnamefont{and}
  \bibinfo{author}{\bibfnamefont{A.~L.} \bibnamefont{Maroto}},
  \bibinfo{journal}{JCAP} \textbf{\bibinfo{volume}{0903}}, \bibinfo{pages}{016}
  (\bibinfo{year}{2009}{\natexlab{b}}), \eprint{0811.0566}.

\bibitem[{\citenamefont{Jimenez et~al.}(2009)\citenamefont{Jimenez, Koivisto,
  Maroto, and Mota}}]{Jimenez:2009sv}
\bibinfo{author}{\bibfnamefont{J.~B.} \bibnamefont{Jimenez}},
  \bibinfo{author}{\bibfnamefont{T.~S.} \bibnamefont{Koivisto}},
  \bibinfo{author}{\bibfnamefont{A.~L.} \bibnamefont{Maroto}},
  \bibnamefont{and} \bibinfo{author}{\bibfnamefont{D.~F.} \bibnamefont{Mota}}
  (\bibinfo{year}{2009}), \eprint{0907.3648}.

\bibitem[{\citenamefont{Novello and Salim}(1979)}]{Novello:1979ik}
\bibinfo{author}{\bibfnamefont{M.}~\bibnamefont{Novello}} \bibnamefont{and}
  \bibinfo{author}{\bibfnamefont{J.~M.} \bibnamefont{Salim}},
  \bibinfo{journal}{Phys. Rev.} \textbf{\bibinfo{volume}{D20}},
  \bibinfo{pages}{377} (\bibinfo{year}{1979}).

\bibitem[{\citenamefont{Novello et~al.}(2007)\citenamefont{Novello, Goulart,
  Salim, and Perez~Bergliaffa}}]{Novello:2006ng}
\bibinfo{author}{\bibfnamefont{M.}~\bibnamefont{Novello}},
  \bibinfo{author}{\bibfnamefont{E.}~\bibnamefont{Goulart}},
  \bibinfo{author}{\bibfnamefont{J.~M.} \bibnamefont{Salim}}, \bibnamefont{and}
  \bibinfo{author}{\bibfnamefont{S.~E.} \bibnamefont{Perez~Bergliaffa}},
  \bibinfo{journal}{Class. Quant. Grav.} \textbf{\bibinfo{volume}{24}},
  \bibinfo{pages}{3021} (\bibinfo{year}{2007}), \eprint{gr-qc/0610043}.

\bibitem[{\citenamefont{Drummond and Hathrell}(1980)}]{Drummond:1979pp}
\bibinfo{author}{\bibfnamefont{I.~T.} \bibnamefont{Drummond}} \bibnamefont{and}
  \bibinfo{author}{\bibfnamefont{S.~J.} \bibnamefont{Hathrell}},
  \bibinfo{journal}{Phys. Rev.} \textbf{\bibinfo{volume}{D22}},
  \bibinfo{pages}{343} (\bibinfo{year}{1980}).

\bibitem[{\citenamefont{Bamba and Odintsov}(2008)}]{Bamba:2008ja}
\bibinfo{author}{\bibfnamefont{K.}~\bibnamefont{Bamba}} \bibnamefont{and}
  \bibinfo{author}{\bibfnamefont{S.~D.} \bibnamefont{Odintsov}},
  \bibinfo{journal}{JCAP} \textbf{\bibinfo{volume}{0804}}, \bibinfo{pages}{024}
  (\bibinfo{year}{2008}), \eprint{0801.0954}.

\bibitem[{\citenamefont{Sadeghi et~al.}(2009)\citenamefont{Sadeghi, Setare, and
  Banijamali}}]{Sadeghi:2009pu}
\bibinfo{author}{\bibfnamefont{J.}~\bibnamefont{Sadeghi}},
  \bibinfo{author}{\bibfnamefont{M.~R.} \bibnamefont{Setare}},
  \bibnamefont{and}
  \bibinfo{author}{\bibfnamefont{A.}~\bibnamefont{Banijamali}}
  (\bibinfo{year}{2009}), \eprint{0906.0713}.

\bibitem[{\citenamefont{Bamba and Nojiri}(2008)}]{Bamba:2008be}
\bibinfo{author}{\bibfnamefont{K.}~\bibnamefont{Bamba}} \bibnamefont{and}
  \bibinfo{author}{\bibfnamefont{S.}~\bibnamefont{Nojiri}}
  (\bibinfo{year}{2008}), \eprint{0811.0150}.

\bibitem[{\citenamefont{Will and Nordtvedt}(1972)}]{Will:1972zz}
\bibinfo{author}{\bibfnamefont{C.~M.} \bibnamefont{Will}} \bibnamefont{and}
  \bibinfo{author}{\bibfnamefont{J.}~\bibnamefont{Nordtvedt},
  \bibfnamefont{Kenneth}}, \bibinfo{journal}{Astrophys. J.}
  \textbf{\bibinfo{volume}{177}}, \bibinfo{pages}{757} (\bibinfo{year}{1972}).

\bibitem[{\citenamefont{Eguchi et~al.}(1980)\citenamefont{Eguchi, Gilkey, and
  Hanson}}]{Eguchi:1980jx}
\bibinfo{author}{\bibfnamefont{T.}~\bibnamefont{Eguchi}},
  \bibinfo{author}{\bibfnamefont{P.~B.} \bibnamefont{Gilkey}},
  \bibnamefont{and} \bibinfo{author}{\bibfnamefont{A.~J.}
  \bibnamefont{Hanson}}, \bibinfo{journal}{Phys. Rept.}
  \textbf{\bibinfo{volume}{66}}, \bibinfo{pages}{213} (\bibinfo{year}{1980}).

\end{thebibliography}

\end{document}